\documentclass{aa}
\usepackage{upgreek}
\usepackage{graphicx}
\usepackage[varg]{txfonts}
\usepackage{natbib}
\usepackage{float}
\usepackage{soul}
\newlength{\linwx}
\usepackage{color}

\begin{document}

\title{How does the chemical composition of solids influence the formation of planetesimals?}

\author{
Konstantinos Odysseas Xenos \inst{1,2}, Bertram Bitsch \inst{2} and Geoffrey Andama\inst{3,4}
}

\offprints{B. Bitsch,\\ \email{bbitsch@ucc.ie}}

\institute{
MAUCA — Master track in Astrophysics, Université Côte d'Azur \& Observatoire de la Côte d'Azur, Parc Valrose, 06100 Nice, France
\and
Department of Physics, University College Cork, Cork, Ireland
\and
Center for Star and Planet Formation, Globe Institute, University of Copenhagen, Øster Voldgade 5–7, 1350 Copenhagen, Denmark
\and
Department of Physics, Muni University, PO Box 725, Arua, Uganda
}

\abstract{
The formation of planetesimals is a necessary step in the formation of planets. While several mechanisms have been proposed, a local dust-to-gas ratio above unity is a strong requirement to trigger the collapse of pebble clouds into planetesimals. A prime location for this is the water-ice line, where large water-rich pebbles evaporate and release their smaller silicate cores. This enhances the local dust-to-gas ratio due to the different inward drift speeds of large and small pebbles. Previous work suggested that planetesimal formation becomes difficult at overall dust-to-gas ratios below 0.6\%, consistent with the occurrence of close-in super Earths. However, the influence of disc composition on planetesimal formation remains unclear. Observations of stellar abundances show both a decrease and a wide spread in C/O ratios for low-metallicity stars. Using the C/O ratio as a proxy to determine water ice abundance in discs, we use the 1D disc evolution code chemcomp to simulate protoplanetary discs with varying C/O and dust-to-gas ratios over 3 Myr. Planetesimal formation is modeled using conditions based on dust-gas dynamics and pebble fragmentation. Our results confirm that planetesimal formation strongly depends on disc metallicity, with lower metallicity discs forming significantly fewer planetesimals. A lower carbon fraction generally promotes planetesimal formation by increasing water ice, while higher carbon fractions suppress it. The opposite is seen for oxygen: higher oxygen content leads to more efficient planetesimal formation at the same dust-to-gas ratio. We thus predict that planets around low-metallicity stars should be more common when their C/O ratio is low and oxygen is enhanced, a trend that can be tested observationally. Our simulations thus open a pathway to understand if the composition of the planet forming material influences the growth of planets.
}


\authorrunning{Xenos et al.}\titlerunning{Planetesimals at low metallicity}\maketitle

\section{Introduction}
\label{sec:Introduction}

Since the detection of the first planet around a main sequence star \citep{1995Natur.378..355M}, significant progress has been made into characterising the occurrence rates of exoplanets via the host star metallicity. Early studies focused on giant planets \citep{2004A&A...415.1153S, 2005ApJ...622.1102F, J2010}, due to the easier detectability of these planets.

The highly successful Kepler mission opened the window to understand also the occurrence rates of smaller planets as function host star metallicity (e.g. \citealt{2013ApJ...766...81F, 2018AJ....156...24M}). Previous studies seem to indicate that planetary radii are a function of the host star metallicity \citep{2012Natur.486..375B, 2014Natur.509..593B, 2018AJ....155...89P, 2018AJ....156..221N}, where planets with smaller radii (below 2 Earth radii) should exist at a wide range of host star metallicities, while larger planets (above 4-5 Earth radii) are preferably found around stars with super-solar metallicity.

The smaller super-Earths and mini-Neptunes are common in systems with multiple planets (e.g. \citealt{2018AJ....156..254W, 2018AJ....155...48W, 2021ApJ...920L..34M}), where these systems can sometimes harbour planets in mean-motion resonances \citep{2016Natur.533..509M, 2017Natur.542..456G}. The formation of these planets can be explained by inward migration and a consequent breaking of the resonances chains that are build during inward migration \citep{2017MNRAS.470.1750I, 2019arXiv190208772I, 2023A&A...674A.178B, 2024A&A...692A.246B}. Interestingly, observations seem to indicate that compact multi-planet systems are more common around metal-poor stellar hosts \citep{2018ApJ...867L...3B}, which could be explained by less instabilities caused by the absence of giant planets around low metallicity stars (e.g. \citealt{2023A&A...674A.178B}).

However, at even lower host star metallicities, [Fe/H]<-0.5, the occurrence rate of planets drops significantly \citep{2024AJ....168..128B}. This thus introduces an interesting test-bed for planet formation theories: what is the metallicity limit at which planet formation can still operate?

Planet formation requires the formation of - at least some - planetesimals, where the final growth to planets can then happen either via planetesimal or pebble accretion (e.g. \citealt{2023ASPC..534..717D}). It is thus crucial to understand the formation of planetesimals within the low metallicity limit. While detailed simulations of the streaming instability (e.g. \citealt{2007ApJ...662..627J, Johansen2015, 2015A&A...579A..43C, 2016ApJ...822...55S}) give great indications what conditions are necessary to form planetesimals (e.g. particle size, local dust-to-gas ratio), they can not cover the evolution and formation of planetesimals over several 100kyrs due to their complex computational requirements.

While several planetesimal formation mechanisms for 1D models exist in the literature (e.g. \citealt{2017A&A...608A..92D, 2017ApJ...839...16C, 2019ApJ...874...36L}), we focus here on the formation of planetesimals at the water ice line \citep{2017A&A...608A..92D}. The underlying idea is that larger water ice particles (with silicate cores) drift inward very rapidly and once they cross the water ice evaporation front, they release their water into the vapour phase, while their smaller silicate cores move slowly inwards. Due to the velocity difference between large and small grains \citep{1977MNRAS.180...57W, 2008A&A...480..859B}, a build-up of solids develops around the water ice line. This can increase the local dust-to-gas ratio above unity and trigger the formation of planetesimals. In our previous work \citep{2024A&A...683A.118A}, we included the recipe for planetesimal formation by \citet{2017A&A...608A..92D} and found that planetesimal formation below [Fe/H] < -0.5 in smooth disc is very challenging - if possible at all, in line with the observational constraints \citep{2024AJ....168..128B}.

However, a crucial ingredient is still missing in these models: the abundance of the different elements (e.g. Fe, Mg, Si, O, C) that contribute significantly to the growth of planets scale differently with [Fe/H] (e.g. \citealt{2021MNRAS.506..150B}) due to the different galactic production sites \citep{1957RvMP...29..547B}. For example, the $\alpha$-elements are normally enriched compared to iron at low metallicity, indicating that maybe this enrichment compensates for the lower iron abundance \citep{2012A&A...543A..89A}. Consequently, planets orbiting stars with lower iron fraction should thus be also less iron rich \citep{2021Sci...374..330A}. However, in total this process should not influence the formation of planetesimals or planets, as it should not matter if a planet accretes iron rich or iron poor pebbles, as long as there is enough material available to allow the growth of the planet.

On the other hand, as planetesimals could form close to the water ice line, the total water abundance could influence the growth of planetesimals. Observations of stars indicate that the stellar C/O ratio decreases with decreasing [Fe/H], which gives rise to a larger water abundance at low [Fe/H] \citep{2020A&A...633A..10B}. Previous simulations already indicated that a larger water fraction could allow a more efficient formation of giant planets by pebble accretion \citep{2016A&A...590A.101B}. This is caused by the enlarged opacity exterior to the water ice line, which results in a larger aspect ratio that ultimately increases the pebble isolation mass \citep{2014A&A...572A..35L, 2018arXiv180102341B, 2018A&A...615A.110A}. The larger cores allow a more efficient accretion of gas, resulting in a more efficient formation of gas giants compared to discs which are dominated by silicates. However, it is unclear how a change in the water abundance would influence the formation of planetesimals.

In this work, we investigate how the chemical composition of the disc via the C/O ratio - that ultimately sets the water ice abundance - influence the growth of planetesimals. This essentially corresponds to the stage of embryo formation before the effects described by \citet{2016A&A...590A.101B} could take place. While the trend that stars with lower [Fe/H] have a larger water abundance seems clear \citep{2020A&A...633A..10B, 2023arXiv230105034C}, there is a large spread in the data, justifying a more detailed investigation. We will use a simple 1D disc model including pebble growth and drift that allows the formation of planetesimals at the water ice line as in our previous work \citep{2024A&A...683A.118A}. We now investigate a variation of the C/O ratio in the disc at the different metallicities. This allows us to probe how the water abundance influences the formation of planetesimals even if the overall metallicity is the same. We focus here in particular on systems with sub-solar metallicities to disentangle the importance of the water fraction for the formation of planetesimals in the limits where planet formation might not be efficient any more.

Our paper is structure as follows. In section~\ref{sec:methods} we discuss our methods and present the corresponding results at high and low C/O ratios in section~\ref{sec:results}. We then discuss the implications of our results in section~\ref{sec:disc} and summarise in section~\ref{sec:summary}.

\section{Methods}
\label{sec:methods}

We utilize \texttt{chemcomp}, a 1 dimensional code simulating a protoplanetary disc's evolution, from dust growth and accretion to planetesimals to planet formation and migration \citep{2021A&A...654A..71S}. We briefly state how we model the disc's evolution, and how we include the formation of planetesimals. We then discuss our chemical model before discussing the parameters we use in our simulations.

\subsection{Disc evolution and planetesimal formation}

\texttt{chemcomp} models the 1 dimensional disc using a viscous evolution model \citep{1973A&A....24..337S}, with the viscocity given by
\begin{equation}
    \nu = \alpha_t \frac{c_s^2}{\Omega_K}
\end{equation}
with $c_s$ the isothermal sound speed in the disc's medium, $\Omega_k = \sqrt{\frac{G M_*}{r^3}}$ the Keplerian angular frequency and $\alpha_t$ a dimensionless factor describing the level of turbulence in the disc. As the disc temperature profile remains the same in time for the duration of the simulations presented below, the sound velocity and by extension the viscosity profiles remain constant in time in each simulation. Varying viscosities are thus probed by variations in the $\alpha_{\rm t}$ parameter.

The temperature profile in our disc is calculate via an equilibrium between viscous and stellar heating with radiative cooling \citep{2021A&A...654A..71S} at the beginning of each simulations. For simplicity we do not evolve the temperature of the disc in time. The disc's surface density is calculated from the initial disc mass and radius in a self-consistent manner corresponding to the steady-state solution of viscous discs. Consequently the disc harbours a radial pressure gradient, which in turn results in a sub-Keplerian velocity of the gas.

The growth and evolution of the dust particles is modelled via the two-population approach \citep{2012A&A...539A.148B}, where small dust grains grow to mm-cm sized pebbles. These particles want to orbit at a Keplerian velocity and thus feel a head wind from the gas, which results in efficient inward movement of the mm-cm sized pebbles \citep{1977MNRAS.180...57W, 2008A&A...480..859B}. The exact velocity of the pebbles depends on the Stokes number of the particles, which essentially describes how well coupled the particles are to the gas. Particles with small Stokes numbers (of the order of $<10^{-4}$) are perfectly coupled to the gas and move with the gas velocity, while large particles decouple from the gas and move inwards much more efficiently. The Stokes number within the Epstein drag regime for spherical particles can be written as
\begin{equation}
    {\rm St} = \frac{a_{\rm p} \rho_s}{\Sigma_g} \frac{\pi}{2}
\end{equation}
with $a_{\rm p}$ and $\rho_s$ the radius and density of the particle respectively and $\Sigma_g$ the surface density of the gaseous environment.

The growth (and thus the drift) of particles is limited by the fragmentation velocity, which essentially describes how fast particles can collide so that they can still stick and grow. Once the velocity of the particles becomes larger, they fragment and the growth process is stopped. The fragmentation velocity is dependent on the composition of the particles and here $u_\text{frag} = 1$ m/s is considered for silicates and $u_\text{frag} = 10$ m/s for water ice \citep{2015ApJ...798...34G}. This result in a large change of the particle sizes at the water ice line, as they depend quadratically on the fragmentation velocity $u_{\rm f}$
\begin{equation}
 \label{eq:afrag}
 a = \frac{2}{3 \pi} \frac{\Sigma_{\rm d}}{\rho_{\rm s} \alpha_{\rm t}} \frac{u_{\rm f}^2}{c_{\rm s}^2} \ .
\end{equation}
Effectively this result in a change of the particle size by a factor of 100 when the fragmentation velocity changes from 10 to 1 m/s at the water ice line.

This process is of great interest in planetesimal formation, as an abrupt switch in the fragmentation velocity occurs at the water evaporation line - so within a very small area within the disc. Suddenly particles reduce in size as they fragment and they couple more to the gas as their Stokes number goes down, forming a traffic jam of solid particles in the area. This traffic jam around the water ice line can lead to the formation of planetesimals (e.g. \citealt{2017A&A...608A..92D}). They give a simple estimate for the formation of planetesimals
\begin{equation}\label{recipe}
    \dot{\Sigma}_\text{pla} = \zeta \cdot \Sigma_{\text{d}} \cdot \Omega_k
\end{equation}
with $\zeta$ the efficiency of the formation, equal to $10^{-3}$, $\Sigma_{\text{d}}$ the surface density of the dust and $\Omega_k$ the keplerian frequency. Planetesimal formation occurs when the condition layed out by \cite{2021ApJ...919..107L} are met:
\begin{enumerate}
    \item The pebble Stokes number is St $\ge 10^{-3}$
    \item The mid plane dust-to-gas ratio is $\epsilon \ge 1$
\end{enumerate}
Here, the mid plane dust-to-gas ratio is computed in the following manner:
\begin{equation}
    \epsilon = \frac{\Sigma_\text{d}}{\Sigma_\text{g}} \sqrt{\frac{\alpha_z + \text{St}}{\alpha_z}}
\end{equation}
where $\Sigma_\text{d}$ and $\Sigma_\text{g}$ is the surface density of the dust and gas respectively, St the Stokes number and $\alpha_z$ the vertical settling parameter, a unit less factor which describes the vertical component of the disc turbulence which is kept constant for all simulations and equal to $10^{-4}$.

\subsection{Chemical model}

The solar elemental abundances \citep{2009ARA&A..47..481A} are used in the disc, listed in Tab.~\ref{elabu}, and the individual elements are then distributed in the different molecules as listed in Tab.~\ref{molecules}. We use here a reduced chemical partitioning model compared to \citealt{2021A&A...654A..71S}, as the stellar abundances used as motivation for our model do not contain N, Na, Ti, V, Al, and S. The large refractory carbon fraction represents the organics in our model, which are thought to hold the largest contributions \citep{1977ApJ...217..425M, 2004ApJS..152..211Z, 2017A&A...606A..16G}.

\begin{table}
\caption{Elemental abundances of the modelled disc.}
\centering
\begin{tabular}{cc}
\hline
Element & Abundance \\ \hline
O/H & $4.90 \times 10^{-4}$  \\
C/H & $2.69 \times 10^{-4}$ \\
Mg/H & $3.98 \times 10^{-5}$ \\
Si/H & $3.24 \times 10^{-5}$ \\
Fe/H & $3.16 \times 10^{-5}$ \\
S/H & $1.30 \times 10^{-5}$ \\ \hline
\end{tabular}
\tablefoot{The stellar abundances are taken from \citet{2009ARA&A..47..481A}.}
\label{elabu}
\end{table}

\begin{table}
\caption{Distribution of the elements into molecules within the modelled disc.}
\centering
\begin{tabular}{ccc}
\hline
Molecule & T$_\text{cond}$ (K) & Volume mixing ratio \\ \hline
CO & 20 & 0.2 $\times$ C/H \\
CH$_4$ & 30 & 0.1 $\times$ C/H \\
CO$_2$ & 70 & 0.1 $\times$ C/H \\
H$_2$O & 150 & \begin{tabular}[c]{@{}c@{}}O/H - (CO/H + 2 $\times$ CO$_2$/H + \\ 4 $\times$ Mg$_2$SiO$_4$/H +3 $\times$ MgSiO$_3$/H )\end{tabular} \\
C & 631 & 0.6 $\times$ C/H \\
Mg$_2$SiO$_4$ & 1354 & Mg/H - Si/H \\
MgSiO$_3$ & 1500 & Mg/H - 2 $\times$ (Mg/H - Si/H) \\
Fe & 1357 & Fe/H \\ \hline
\end{tabular}
\tablefoot{Volume mixing ratios of each species are based on the works of \citet{2014ApJ...794L..12M}, \citet{2020A&A...633A..10B}, and \citet{2021A&A...654A..71S}, and their condensation temperatures are from \citet{2003ApJ...591.1220L}. The used methane fraction corresponds roughly to observations \citep{2004ApJS..151...35G, 2011ARA&A..49..471M}.
}
\label{molecules}
\end{table}

The C/O ratio of the disc is derived using the abundances in Tab. \ref{elabu} and the scaling of \cite{2020A&A...633A..10B} which originate from the mean values extracted from the GALAH (Galactic Archaeology with HERMES) survey \citep{2018MNRAS.478.4513B}. In order to explore the parameter space for higher and lower C/O ratios, a C/O value $\pm 0.25$ from the mean C/O was selected at each [Fe/H]. The C/O ratio of the disc can be tuned by carefully selecting the [C/H] or [O/H] abundance:
\begin{align}
    \text{[C/H]} &= \log_{10}\left(\frac{\text{O/H}}{\text{C/H}} \right) + \log_{10}(\text{C/O}) + \text{[O/H]} \\
    \text{[O/H]} &= \log_{10}\left(\frac{\text{C/H}}{\text{O/H}} \right) - \log_{10}(\text{C/O}) + \text{[C/H]}
\end{align}
where O/H and C/H are the elemental abundances of O and C, listed in Tab. \ref{elabu}, C/O is the number ratio of C over O and [O/H] and [C/H] are the abundances of O and H in dex notation.

Our aim is to vary the C/O ratio within the disc, which can be varied by either changing the overall carbon or overall oxygen abundances. This then results in different [C/H] and [O/H] values for the same C/O. The corresponding [C/H] and [O/H] abundances calculated for the C/O ratios of interest ($\pm 0.25$ from the mean value) are listed in Tab.~\ref{getCH}.

\begin{table}[H]
\caption{Dust-to-gas ratios tested for varying C/O ratios.}
\centering
\begin{tabular}{ccccc}
\hline
{[}Fe/H{]} & Dust to gas ratio & C/O ratio & {[}C/H{]} & {[}O/H{]} \\ \hline
0.0 & 0.016 & 0.30, 0.80 & -0.23, 0.19 & 0.29, -0.13 \\
-0.2 & 0.011 & 0.23, 0.73 & -0.46, 0.04 & 0.23, -0.26 \\
-0.4 & 0.008 & 0.19, 0.69 & -0.65, -0.09 & 0.17, -0.39 \\
 & 0.007 & 0.19, 0.69 & -0.65, -0.09 & 0.17, -0.39 \\
 & 0.006 & 0.19, 0.69 & -0.65, -0.09 & 0.17, -0.39 \\
 & 0.005 & 0.19, 0.69 & -0.65, -0.09 & 0.17, -0.39 \\
 & 0.004 & 0.19, 0.69 & -0.65, -0.09 & 0.17, -0.39 \\ \hline
\end{tabular}
\tablefoot{We only list [Fe/H] down to -0.4, as this corresponds to the chemical abundances study by \citet{2020A&A...633A..10B} and as the data becomes more sparse at lower [Fe/H].}
\label{getCH}
\end{table}

As we only have the data for reliable C/O ratios down to [Fe/H]=-0.4, we adopt the same C/O ratios (and derivations thereof) for even lower dust-to-gas ratios. A change in the C/O ratio will result in a change of the overall water abundance in the disc. If there is less carbon available, more oxygen will be free to form water, while a larger C/O ratio will result in less water, as more oxygen is bound in CO and CO$_2$. We note that also variations of the Mg and Si abundances will change the disc's water content, however changes in the carbon and oxygen fraction will have a larger influence due the larger amount of oxygen and carbon (see table~\ref{elabu} and \citealt{2023arXiv230105034C}).

In addition, the two different methods to change the C/O ratio (either increasing the carbon or oxygen abundance) have different consequences for the amount of water as well. If the overall oxygen abundance is increased (to mimic a lower C/O ratio), the overall mass of all other elements has to be reduced for a constant dust-to-gas ratio. Consequently, a low C/O ratio realised by a high oxygen content will feature more water, compared to a low C/O ratio that is realised by decreasing carbon, as this increases the amount of all other solids (and not only water), which also bind oxygen (see table~\ref{molecules}). We show the water ice fractions in our model in more detail in appendix~\ref{app:water}.

\subsection{Disc parameters}

In our model, the stellar mass, disc mass and disc radius were constant throughout, as we want to investigate the influence of the chemical composition on planetesimal formation. In particular we use here a solar mass star with a disc that corresponds to a 10\% solar mass disc, which is on the heavier side of protoplanetary discs. We've chosen this value as we aim to test the limits of planetesimal formation at low metallicity. Our previous work indicated that the efficiency for planetesimal formation (defined as how much mass of the dust is converted into planetesimals) is slightly higher in higher mass discs \citep{2024A&A...683A.118A} and can be up to 50\%, depending also on the disc's viscosity. The turbulent viscosity of the disc is varied between $10^{-3}$ and $10^{-4}$ and the vertical settling parameter of the disc is kept constant at $\alpha_{\rm z}=10^{-4}$ \citep{Pinilla_2021}.

The stellar metallicity, which is assumed to be also that of the disc, is set at [Fe/H] = 0 and two sub solar values, -0.2 and -0.4 before we investigate even lower dust-to-gas ratios. The remaining elemental abundances are taken from the scaling law described in \cite{2020A&A...633A..10B}. In this model, only Fe, O, C, Si, S and Mg are assumed to exist in substantial quantities, while we neglect other species. With the elemental abundances under consideration in the model, the initial dust-to-gas ratio (DTG) of the disc is then given by
\begin{equation}
    {\rm DTG} = \sum{\text{X/H}} \cdot \mu_x \cdot 10^{\text{[X/H]}}
    \label{eqn:dtg}
\end{equation}
with X/H the initial abundance of element X \citep{2009ARA&A..47..481A}, $\mu_x$ the atomic mass of element X and [X/H] the abundance of element X \citep{2020A&A...633A..10B} in dex units. This gives a dust-to-gas ratio of $\approx 1.6$\% for [Fe/H]=0.0.

In addition to the dust-to-gas ratios computed in the previous manner, four lower DTG values were chosen to model very low DTG ratio discs and explore the lower limits of planetesimal formation. The chemical abundances and C/O ratio selected in these cases are identical to the [Fe/H] = -0.4 case, in accordance with the trends from \cite{2020A&A...633A..10B}. We list our chosen disc parameters in table~\ref{tab:parameters}.

\begin{table}[t]
\caption{Disc parameters used in this work}
\centering
\begin{tabular}{ccc}
\hline
Parameter & Symbol & Value \\ \hline
Stellar mass & $M_*$ & 1 $M_\odot$ \\
Disk mass & $M_0$ & 0.1 $M_\odot$ \\
Disk radius & $R_0$ & 100 AU \\
Turbulent viscosity & $a_t$ & $10^{-4}$, $10^{-3}$ \\
Vertical settling & $a_z$ & $10^{-4}$ \\
Dust to gas ratio & DTG & \begin{tabular}[c]{c}0.001, 0.002, 0.003,\\ 0,004, 0.005, 0.006, 0.007, \\ 0.0086, 0.011, 0.016\end{tabular} \\
Stellar metallicity & {[}Fe/H{]} & -0.4, -0.2, 0.0 \\ \hline
\end{tabular}
\label{tab:parameters}
\end{table}

\section{Planetesimal formation}
\label{sec:results}

In this section, we discuss the results of our simulations. We will focus here mainly on the result at overall low metallicity ([Fe/H]=-0.4 corresponding to a dust-to-gas ratio of 0.8\%) and present our results for solar metallicity in appendix~\ref{app:solar}. We first vary the C/O ratio by changing the carbon abundance and then vary the C/O ratio by varying the oxygen abundance.

\subsection{[Fe/H]=-0.4}

\begin{figure}
  \centering
  \includegraphics[width=\hsize]{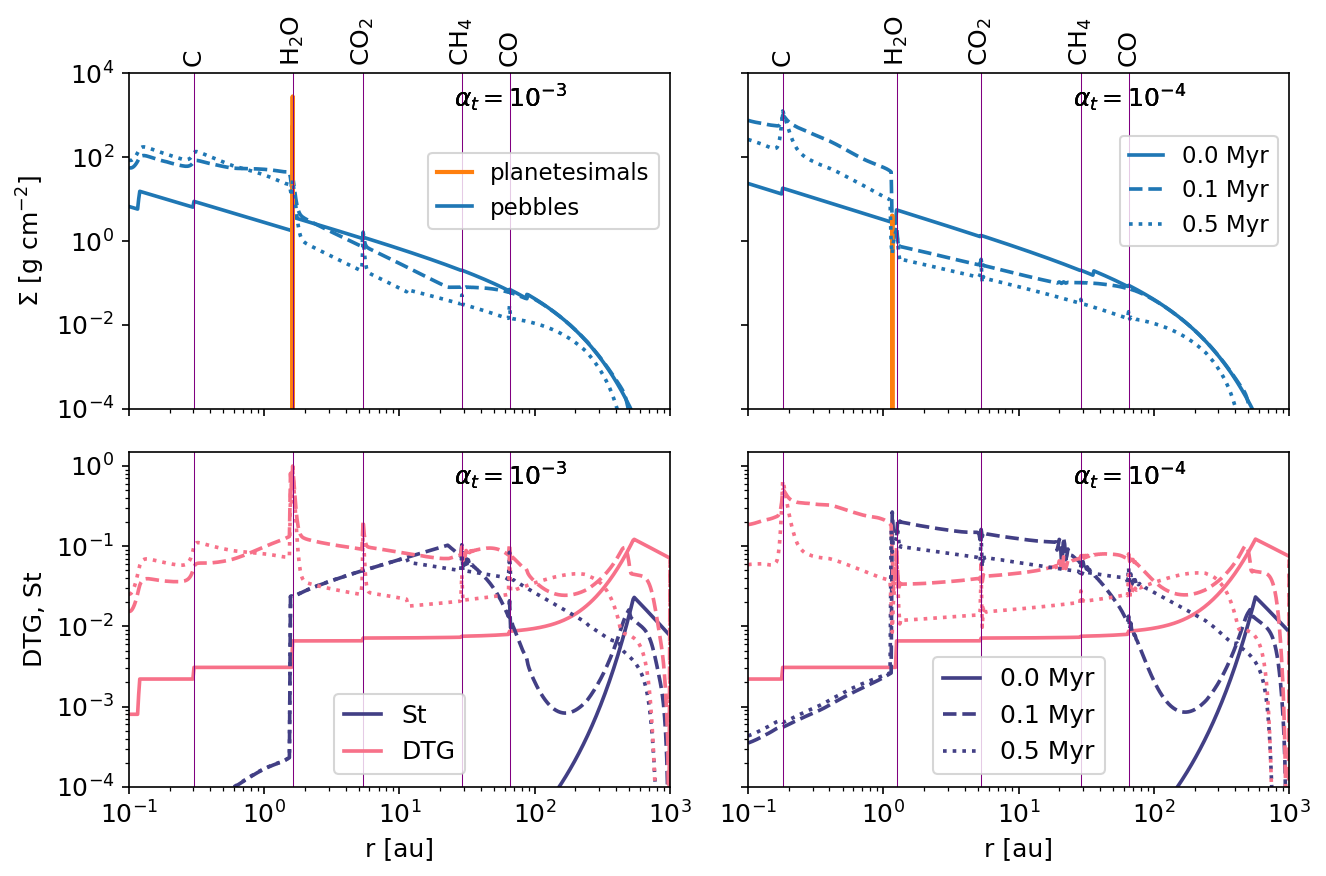}
      \caption{Solid surface density of pebbles and planetesimals (top) in a disc with [Fe/H] = -0.4 and the median C/O ratio for $\alpha=10^{-3}$ (left) and $\alpha=10^{-4}$ (right) as function of time. The bottom panels depict the Stokes numbers of the pebbles as well as the mid-plane dust-to-gas ratio. Initially, the particles are micro meter sized, with Stokes numbers below $10^{-4}$ interior of around 100 AU.}
         \label{fig:neg04COmed}
  \end{figure}

We show in Fig.~\ref{fig:neg04COmed} the time evolution of the pebble and planetesimal surface density as well as of the Stokes number and mid plane dust-to-gas ratio in discs with different viscosities for our median C/O ratio. We focus here mostly on the first few 100kyrs, where planetesimal formation is efficient in our model.

The initial dust-to-gas ratio is below 1\% at the water ice line due to our initial conditions. Initially the discs consist of micro meter sized grains, with low Stokes numbers (bottom of Fig.~\ref{fig:neg04COmed}), but they soon start to grow via coagulation from the inside out. Once the dust growth to mm-cm sized pebbles, their Stokes number increases to the order of $10^{-3}-10^{-2}$ and drift inward efficiently. However, once they cross the water ice line, their size changes by a factor of 100 due to the changes in the fragmentation velocity from 10 m/s to 1 m/s (eq.~\ref{eq:afrag}). Consequently a pile up of material is generated at the water ice line, increasing the mid plane dust-to-gas ratio above unity, allowing the formation of planetesimals at that position.

In time, also the dust in the outer regions of the disc have grown to pebble sizes and drifted inwards. As a consequence, the pebble flux reduces in time. Once the pebble flux is reduced, the dust-to-gas ratio in the mid plane regions of the disc decreases below unity. Consequently planetesimal formation stops.

This directly implies that the formation efficiency of planetesimals at the water ice line is regulated by the efficiency of inward drift of pebbles. The surface density of the formed planetesimals in Fig.~\ref{fig:neg04COmed} is larger for the simulations with higher viscosity. This is caused by the fact that the pebble size in this case are smaller (eq.~\ref{eq:afrag}) and thus drift inwards slower. Consequently, a slightly lower pebble flux (compared to lower viscosities) can be maintained for longer (see also \citealt{2023A&A...679A..11B}). In combination with the low vertical turbulence (fixed at $\alpha_{\rm z}=10^{-4}$), the pebble flux is still large enough to generate a mid-plane dust-to-gas ratio above unity, resulting in the fact that planetesimal formation can happen for a longer period of time. We thus observe a more efficient planetesimal formation at the water ice line at higher viscosity, in line with \citet{2024A&A...683A.118A}.

 \begin{figure*}
  \centering
  \includegraphics[width=0.475\hsize]{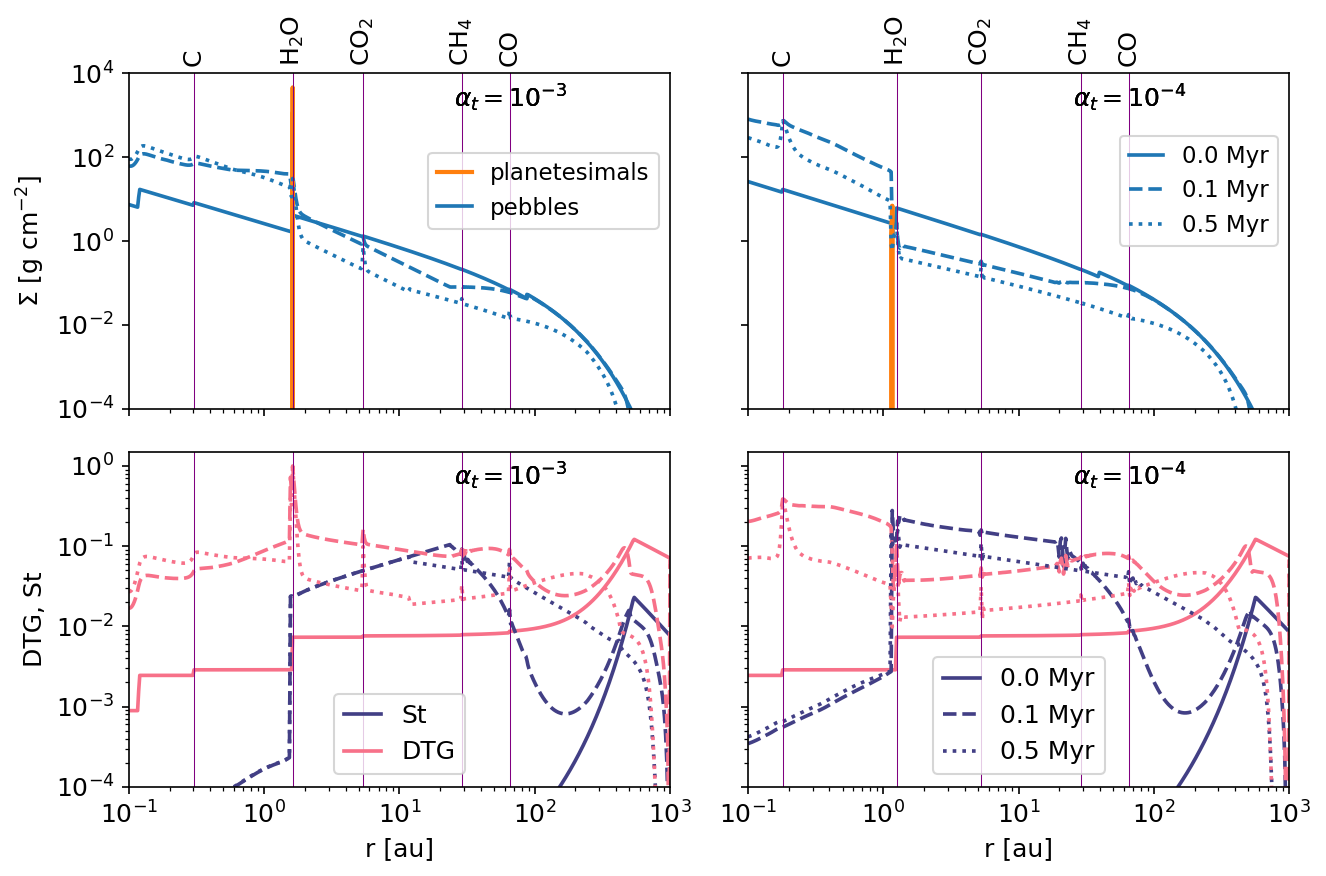} \quad
  \includegraphics[width=0.475\hsize]{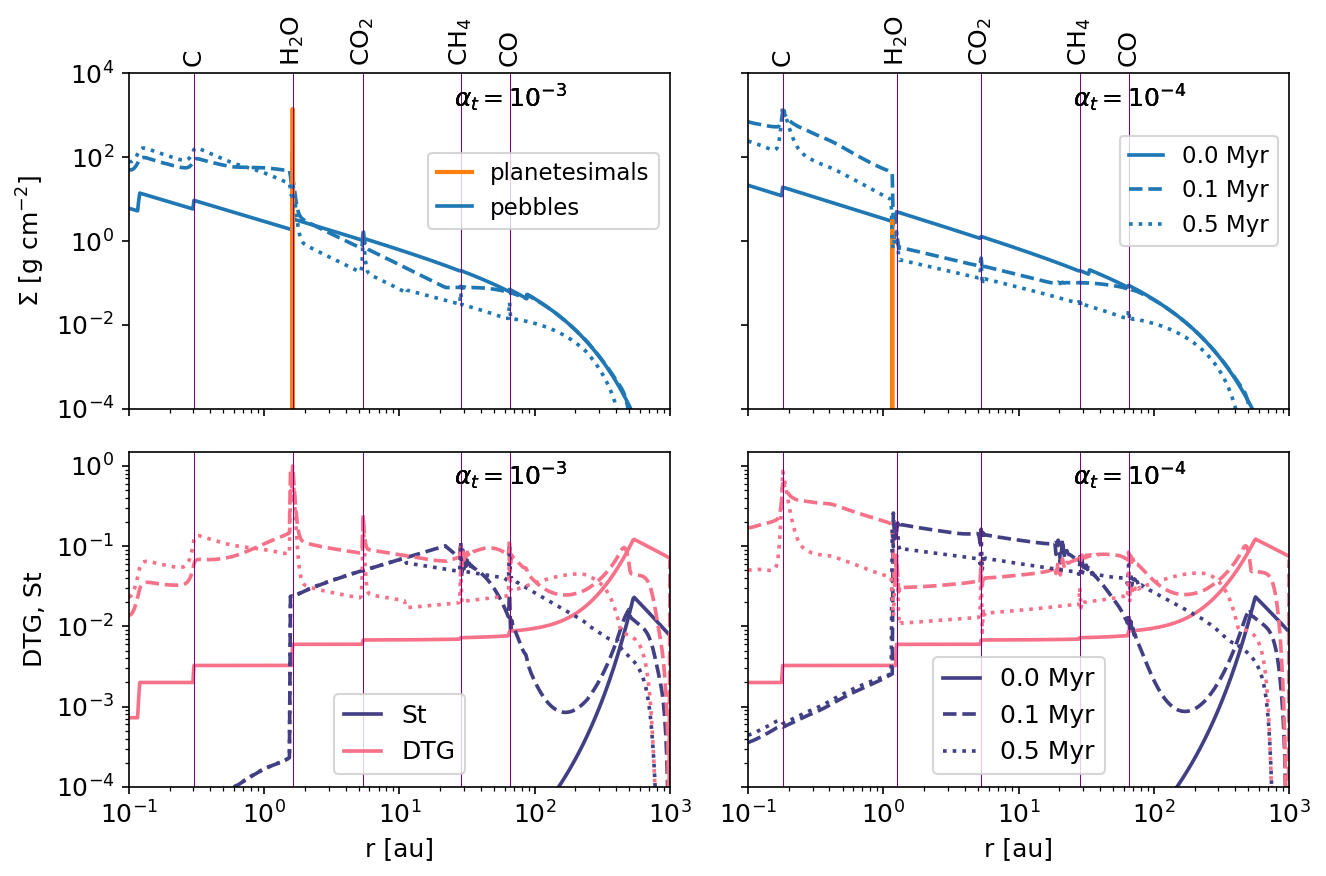}
      \caption{Same as Fig.~\ref{fig:neg04COmed}, but for the low C/O ratio (left) and the high C/O ratio (right).}
         \label{fig:neg04COhighlow}
  \end{figure*}

In Fig.~\ref{fig:neg04COhighlow} we show the evolution of the pebbles and planetesimals for discs with lower C/O ratio (left) and higher C/O ratio (right). We note that we use here the same [Fe/H] value (and thus the same dust-to-gas ratio) as in Fig.~\ref{fig:neg04COmed}. In principle we observe the same phenomenon as for the simulations with the median C/O ratio: the pebbles drift inwards, the dust-to-gas ratio at the water ice line increases above unity and planetesimals can form. Once the pebble flux decreases, planetesimal formation stops. Independently of the C/O ratio, we see that planetesimal formation is more efficient at higher viscosity, caused by the slightly smaller pebbles that can maintain the pebble flux for a longer time resulting in a longer time for planetesimal formation.

A lower C/O ratio results in a larger water content compared to the higher C/O ratio \citep{2020A&A...633A..10B}. Consequently, our simulations show a larger surface density in planetesimals at the water ice line for the simulations with lower C/O ratio.

\begin{figure*}
  \centering
  \includegraphics[width=\hsize]{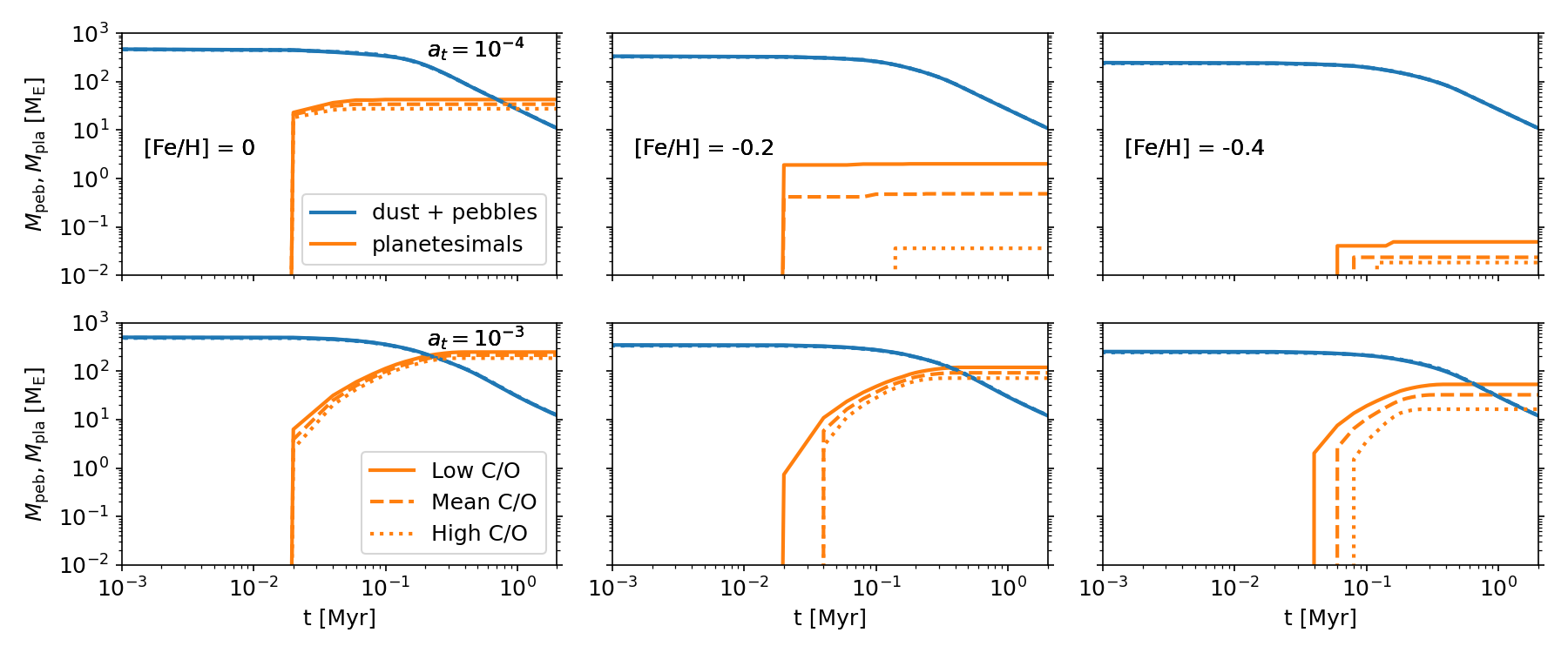}
      \caption{Pebble and planetesimal mass evolution for discs with different [Fe/H] values (left to right) and different viscosities (top to bottom). The different line styles correspond to different C/O ratios, which influence the disc's water content and consequently planetesimal formation.}
         \label{fig:massesFeH}
  \end{figure*}

We show in Fig.~\ref{fig:massesFeH} the mass evolution of dust and pebbles as well as of the planetesimals for discs with [Fe/H]=0.0, -0.2, and -0.4. As in previous works by \citet{2024A&A...683A.118A} and as discussed above, the simulations with larger viscosity form more planetesimals at all [Fe/H] values, as planetesimal formation can be maintained for a longer period of time.

In addition, the simulations with low C/O ratio form planetesimals more efficiently at the same total dust-to-gas ratio. This is, as described above, caused by the larger water fraction at lower C/O ratio. We note that a larger C/O ratio results in a larger CO and CO$_2$ fraction. However, this increase in CO and CO$_2$ is not sufficient to trigger planetesimal formation in the outer disc, as no pebble pile up can be achieved. The large pebble pile-up around the water ice line is caused not only by the water fraction but also by the change in the grain fragmentation velocity from 10 to 1m/s, reducing the grain sizes by a factor of 100. This results in a traffic jam of dust at the water ice line, required to reach local solid-to-gas ratios of unity that allow the formation of planetsimals. In order to achieve a large enough pile up of material at the CO and CO$_2$ snow lines, super-solar metallicities are required (see \citealt{2024A&A...683A.118A}).

\subsection{Lower dust-to-gas ratios}

\begin{figure*}
  \centering
  \includegraphics[width=\hsize]{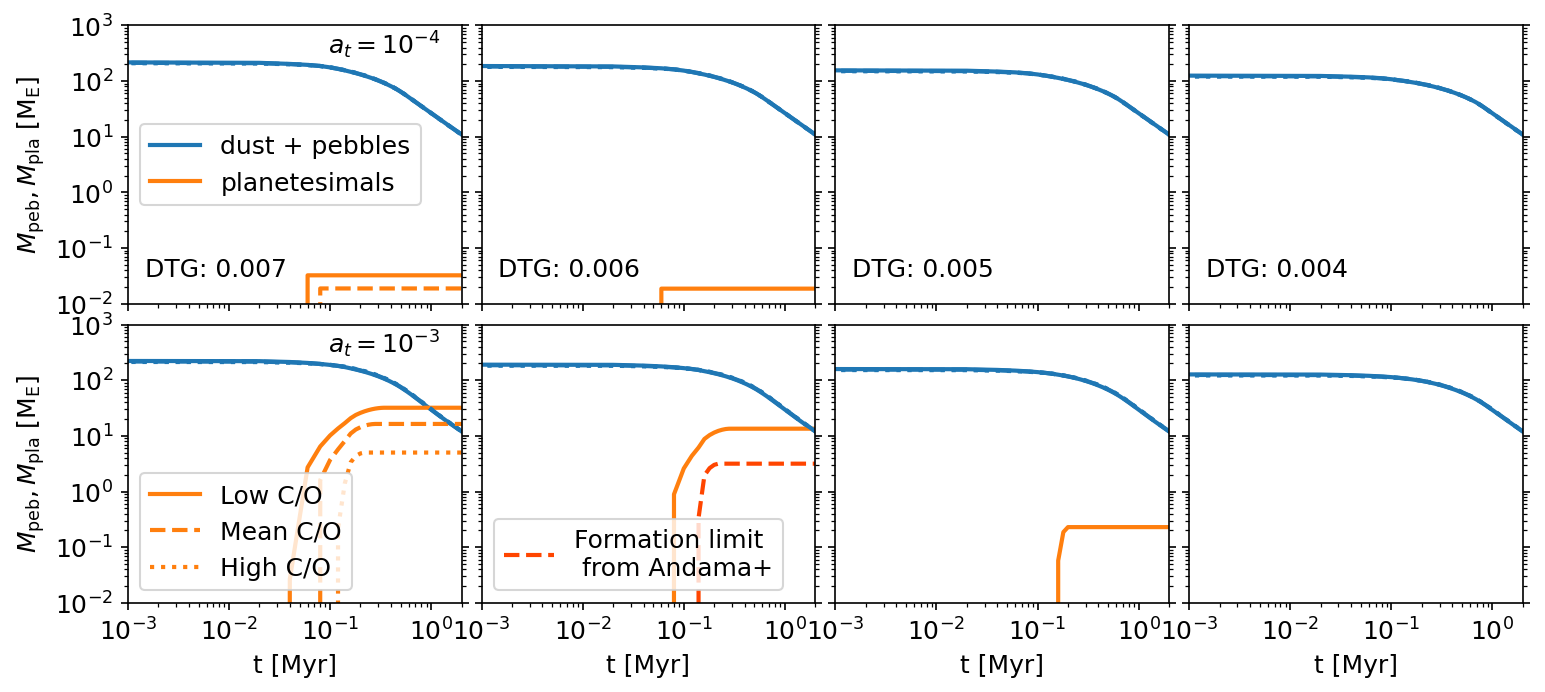}
      \caption{Same as Fig.~\ref{fig:massesFeH}, but for even lower dust-to-gas ratios, where the limit of no planetesimal formation is reached, depending on the initial C/O ratio and initial dust-to-gas ratio.}
         \label{fig:masseslowFeH}
  \end{figure*}

In Fig.~\ref{fig:masseslowFeH} we show the time evolution of the dust and pebbles as well as of the formed planetesimals for simulations with even lower dust-to-gas ratios. The simulations with low viscosity ($\alpha_{\rm t}=10^{-4}$) only produce a very tiny amount of planetesimals at dust-to-gas ratios of 0.7\% and 0.6\%, and nothing below these dust-to-gas ratios, independently of the C/O ratio of the material. Clearly, planet formation by pure planetesimal accretion will be troublesome at dust-to-gas ratios of 0.7\% and 0.6\%. On the other hand, a formation of super-Earths assisted by pebble accretion and giant impacts might be possible, even at these low pebble fluxes (e.g. \citealt{2019A&A...627A..83L}).

On the other hand, as discussed above, the higher viscosity ($\alpha_{\rm t}=10^{-3}$) allows to maintain the pebble flux (and a mid-plane dust-to-gas ratio above unity) for longer and thus allows a more efficient planetesimal formation at the water ice line. In particular for a dust-to-gas ratio of 0.7\% and 0.6\%, our simulations predict a ring of planetesimals at the water ice line of more than 5-10 Earth masses, which is enough to form super-Earths, even with pure planetesimal accretion \citep{2022NatAs...6..357I, 2023NatAs.tmp...10B, 2025ApJ...979L..23S}. Our simulations also show clear dependency on the amount of planetesimals that are form as function of the C/O ratio. At a dust-to-gas ratio of 0.6\%, the simulations with high C/O ratio fail to form any planetesimals, due to the low water content that prevents planetesimal formation at the water ice line.

At a dust-to-gas ratio of 0.5\% our simulations only show a small amount of planetesimals that are formed and only at low C/O ratios, where the water content is larger compared to the simulations with higher C/O ratios. The amount of planetesimals formed at this dust-to-gas ratio would be too low to form planets by planetesimal accretion, but planets could also grow by pebble accretion and giant impacts during or after the gas disc phase.

\begin{figure*}
  \centering
  \includegraphics[width=\hsize]{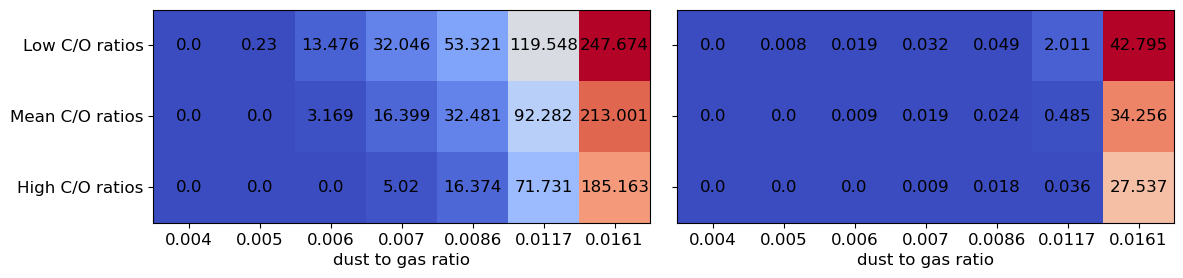}
      \caption{Planetesimal mass (in Earth masses) that is formed in discs with different initial dust-to-gas ratios as well as different C/O ratios. The change of the C/O ratio is achieved by changing the carbon grain fraction. The left panel depicts simulations with high viscosity, while the right panel shows simulations with low viscosity.}
         \label{fig:massesfull}
  \end{figure*}

We show in Fig.~\ref{fig:massesfull} the total masses of planetesimals that are formed in our simulations. As discussed above, the final mass in planetesimals is essentially negligible below solar metallicity in low turbulence discs, while discs with higher turbulence can form planetesimals at lower dust-to-gas ratios. The trend that planetesimals form more efficiently with lower C/O ratio is clearly visible.

\subsection{Changing the total oxygen abundance}

The C/O ratio within the disc can also be changed by changing the oxygen abundances. Above we changed the C/O ratio by changing the carbon fraction, while we now change the oxygen abundance in the disc to manipulate the C/O ratio. This has consequences for the water content within the disc. Low C/O ratios imply an increase in the oxygen abundance. The total abundance of all other elements then has to decrease as the dust-to-gas ratio remains fixed. Consequently, more water is available, influencing the formation efficiency of planetesimals at the water ice line.

\begin{figure*}
  \centering
  \includegraphics[width=\hsize]{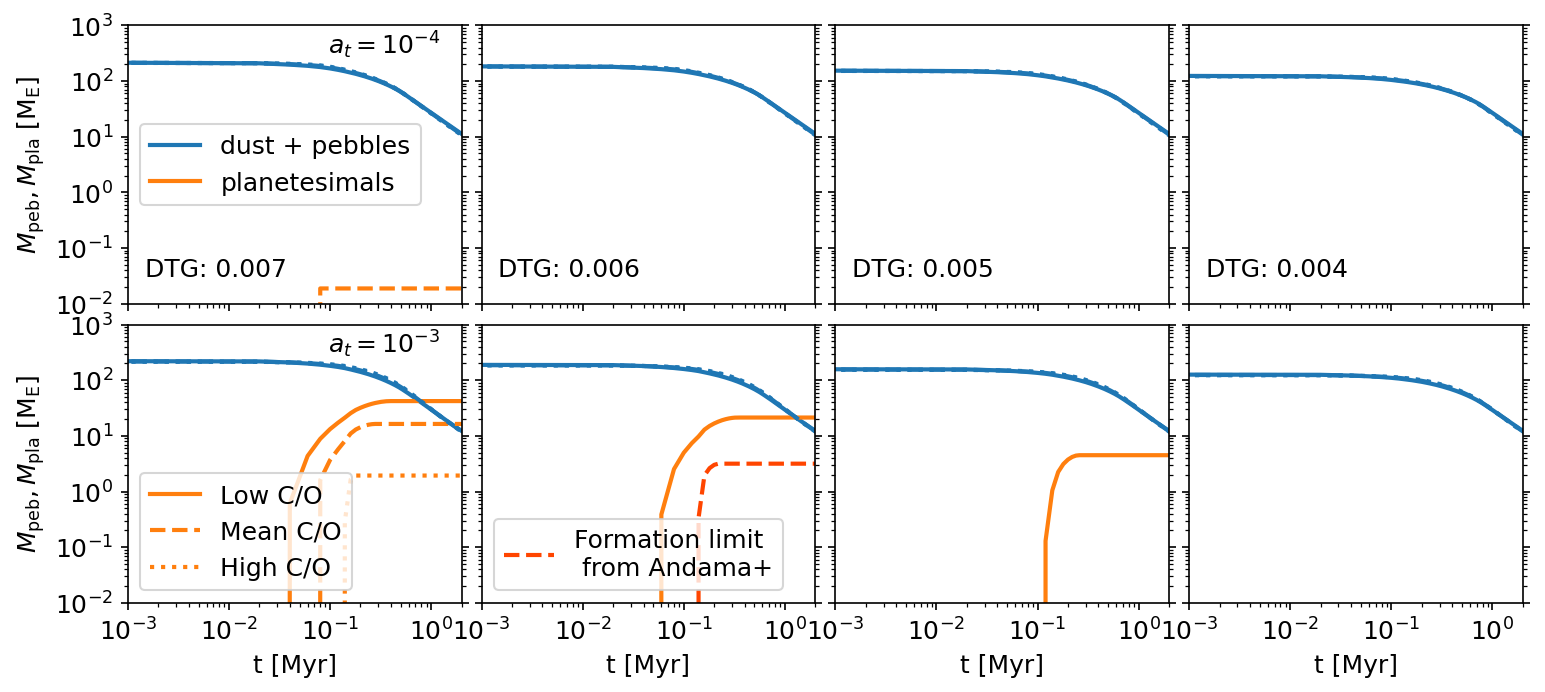}
      \caption{Same as Fig.~\ref{fig:masseslowFeH}, but for discs where the C/O ratio is varied by changing the oxygen abundance.}
         \label{fig:masseslowFeHoh}
  \end{figure*}

\begin{figure*}
  \centering
  \includegraphics[width=\hsize]{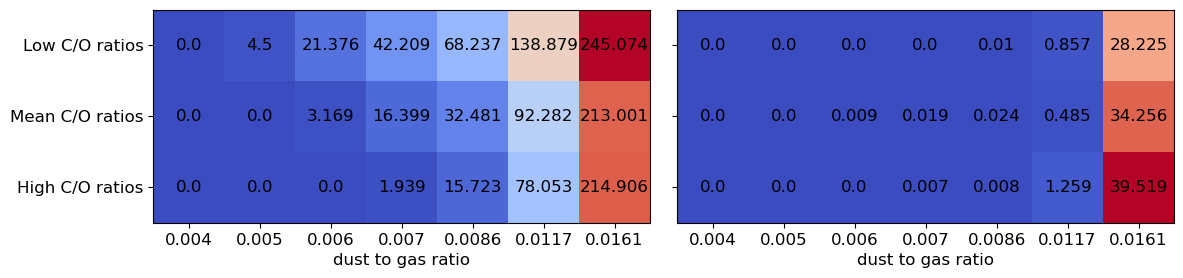}
      \caption{Same as Fig.~\ref{fig:massesfull}, but for discs where the C/O ratio is varied by changing the oxygen abundance.}
         \label{fig:massesfulloh}
  \end{figure*}

In Fig.~\ref{fig:masseslowFeHoh} we show the dust and pebble as well as planetesimal mass as function of time for low dust-to-gas ratios for different viscosities. In Fig.~\ref{fig:massesfulloh} we show the total planetesimal mass that is formed in our simulations.

Our simulations show the same trend as before, that discs with low turbulence do not form planetesimals, while discs with higher levels of turbulence form planetesimals more efficiently, especially if the C/O ratio is lower. A notable difference compared to the previous simulations is that at a dust-to-gas ratio of 0.5\% several Earth masses of planetesimals formed at low C/O ratio, in contrast to the simulations with high C/O ratio. These masses could be high enough to form planets even by planetesimal accretion alone \citep{2022NatAs...6..357I, 2023NatAs.tmp...10B, 2025ApJ...979L..23S}. Consequently the trend that planetesimal and thus planet formation at low metallicity is influenced by the disc's C/O ratio (and thus water content) holds also in this scenario.

We note that the formation efficiency of planetesimals at solar metallicity is also influence by the overall C/O ratio. However, we expect that this will not make a significant difference in the formation of planets, because very large amounts of planetesimals (above a few 10 Earth masses) are formed in any case, allowing the very efficient formation of planets in all cases.

\section{Discussion}
\label{sec:disc}

\subsection{Formation at pressure perturbations}

The size difference between the particles interior and exterior to the water ice line results in an opacity difference at the region, which consequently gives rise to a very steep radial temperature gradient. This could then cause a pressure perturbation in the disc \citep{2021A&A...650A.185M}. Within this pressure perturbation, pebbles could accumulate and the formation of planetesimals could be triggered (e.g. \citealt{2007Natur.448.1022J}). In addition, observations indicate that protoplanetary discs harbour substructures, where particles could accumulate, eventually allowing the formation of planetesimals and planets within these substructures \citep{2020A&A...638A...1M, 2024A&A...686A..78S}. However, it is unclear how these substructures are generated (see e.g. \citealt{2023ASPC..534..423B}) and when they appear in protoplanetary discs.

From a mass budget perspective, even a disc with a dust-to-gas ratio of only 0.5\% harbours around 150 Earth masses of solids, if the disc was initially 10\% the mass of the sun. If all of this pebble mass were to accumulate in one ring, clearly enough planetesimals could be formed to facilitate the formation of planets in this scenario, in contrast to our here presented simulations that feature a smooth disc without any initial perturbations.

This opens the question that if substructures are common and allow the efficient formation of planetesimals and planets, why is there a lower limit in the [Fe/H] at around $\sim -0.5$ for the occurrence rate of planets? Massive discs with even lower dust-to-gas ratios that 0.5\% should harbour enough mass to form planetesimals and planets, if it is all accumulated within one ring. Clearly there are several possible answers to this question, where the obvious answer is that disc are on average much smaller than 10\% of the stellar mass (e.g. \citealt{2025A&A...696A.232G}). Reducing the disc mass to something like the Minimum-Mass-Solar-Nebula \citep{1977Ap&SS..51..153W, 1981PThPS..70...35H}, corresponding to roughly 1\% of the solar mass, reduces the total solid mass to only 15 Earth masses, when considering a 0.5\% metallicity, where even the ring formation scenario runs into troubles. In addition, most close in super-Earths and mini-Neptunes are found around M and K-dwarf stars, which have a lower mass and consequently a lower disc mass and total solid mass budget.

Alternatively, the question remains when the substructures are generated in the protoplanetary discs. Observations seem to indicate that class 0 discs have mostly no detected sub-structures \citep{2023ApJ...951....8O}, while the fraction of discs with sub-structures increases towards class I and class II sources \citep{2024ApJ...973..138H, 2025arXiv250411577H}. In this case though, the majority of the pebbles might have already drifted inwards, leaving only smaller masses of pebbles available to form planetesimals at pressure perturbations.

It is thus clear that this question could only be answered if the initial disc mass function was known. As it stands, it seems that the lack of planets at low [Fe/H] indicates formation within smooth discs, as the limits of planetesimal formation coincide with the limits of the occurrence rates of close-in planets or that the observed pressure perturbations - visible through gaps and rings - form too late to allow an efficient formation of planetesimals and planets.

\subsection{Particle sizes}

Our model predicts that planetesimal formation in disc with $\alpha_{\rm t}=10^{-3}$ is more efficient compared to discs with a lower turbulence. This is related to the fact that the particle sizes in discs with higher turbulence are smaller (eq.~\ref{eq:afrag}) and consequently drift inwards slower, allowing the pebble flux to be maintained for a longer time. This implies that the mid plane dust-to-gas ratio at the water ice line can be above unity for a longer time, giving a longer time during which planetesimals can form. As a result these simulations show a larger planetesimal mass.

However, this is based on a crucial assumption in our simulations: that the vertical level of turbulence, which determines the settling and thus the mid plane dust-to-gas density (eq.~\ref{eqn:dtg}), is low ($\alpha_{\rm z}=10^{-4}$). This assumption is motivated by the study of \citet{Pinilla_2021}, who showed that a low vertical turbulence is needed to explain the observed concentration of large particles in the mid plane of observed protoplanetary discs. If the level of vertical stirring/turbulence would be higher, planetesimal formation will become more difficult or not happen at all, as the mid-plane dust-to-gas ratio does not exceed unity \citep{2017A&A...608A..92D, 2024A&A...683A.118A}. The question about the connection between the different assumed turbulences related to i) vertical settling, ii) disc viscous evolution, and iii) grain fragmentation still remains open.

We note that the crucial parameter for a more efficient planetesimal formation - the particle size - could also be manipulated by different assumptions for the fragmentation velocity of the grains. In our model we assume 10m/s for icy particles and 1m/s for silicate particles, motivated by laboratory experiments \citep{2015ApJ...798...34G}. However, it is still open if water ice particles really have a larger fragmentation velocity compared to silicate particles (e.g. \citealt{2019ApJ...873...58M}). It is clear though, that a small change in the fragmentation velocity of grains could compensate a change in the turbulence regarding the particle size (eq.~\ref{eq:afrag}). Consequently, we could recover the same grain sizes as in our $\alpha_{\rm t}=10^{-3}$ simulations within simulations with $\alpha_{\rm t}=10^{-4}$ by using a fragmentation velocity of $\sqrt{10}$m/s rather than 10m/s. We expect that planetesimal formation would then still be as efficient as in our nominal $\alpha_{\rm t}=10^{-3}$ case. While a detailed study of how the fragmentation velocity impacts the formation of planetesimals is beyond the scope of this study, it is clear that the water abundance in the disc would still influence the formation of efficiency of planetesimals at the water ice line.

In principle CO and CO$_2$ ice could be trapped inside water ice rich pebbles and be transported into the inner disc regions. Recent studies with JWST imply that the trapped CO-to-water fraction is of the order of 15\% \citep{2024ApJ...975..166B}. However, the trapped CO ice would evaporate before the water ice line at around 130K \citep{burke2010}, meaning it would not add to the pile-up of material at the water ice line. In addition, trapped CO ice would not re-condense at the water ice line, due to its lower evaporation front. We thus think these effects would be minimal in our model.

\subsection{Implications for exoplanets and planet formation}

Our simulations indicate that the formation of a ring of planetesimals at the water ice line is possible at low metallicity, in agreement with our previous simulations \citep{2024A&A...683A.118A}. Our simulations, however, do not include self consistently the formation of planetary embryos and thus planets. In the inner region this process could take a few 100kyr \citep{2022A&A...666A..90V}. Only then would we expect significant planet-disc interactions to start, which could cause pressure perturbations. However, at this stage our simulations do not produce any planetesimals any more (see Fig.~\ref{fig:massesFeH}), indicating that the formation of planets would not significantly impact our result, especially at low metallicity where only a small amount of planetesimals is formed in the first place (Fig.~\ref{fig:massesfull}, Fig.~\ref{fig:massesfulloh}). However, we are interested here in the limits of planetesimal formation at low metallicity, where only a tiny amount of planetesimals is formed.

Our simulations predict an significant increase of planetesimals that could be formed at the water ice line in case of low C/O ratios - that indicate a large water ice content (Appendix.~\ref{app:water}). This implies that there could be another layer related to the occurrence rate of planets at low metallicity, which has not been probed before: the overall C/O ratio coupled with a large water content. Our simulations show that planetesimal formation in general is very difficult at overall dust-to-gas ratios of $\le 0.6\%$. However, at these low dust-to-gas ratios, our simulations show a significant increase in planetesimal formation at low C/O ratios, implying large water fractions. We thus expect that exoplanets detected around stars with low [Fe/H] should harbour a low C/O fraction as well and a large oxygen/water fraction. Our simulations thus show that the overall metallicity, which is determined by the summation over all elements, should be used to determine exoplanet occurrence rates in contrast to [Fe/H], which is normally used and does not contain any information about the total dust-to-gas ratio or the water fraction.

Detailed stellar abundance observations of exoplanet host stars at low [Fe/H] could thus be used to test our model. It is important to note that stars with low [Fe/H] are typically enriched in $\alpha$-elements \citep{2021MNRAS.506..150B, 2023arXiv230105034C}, meaning that the overall metallicity has to be determined to allow a comparison to our model. \citet{2021A&A...655A..99D} studied the detailed stellar abundances of 1111 stars, including 152 planet host stars. They find that the stellar C/O ratio of exoplanet host stars (and stars without detected planet) with [Fe/H]<-0.4 are of the order of 0.2-0.4, so within the range of the C/O ratios of our model (table~\ref{getCH}). However, we can not draw a definite conclusion from their sample, as only a handful of planet host stars are within that range, making a statistical analysis impossible. However, the upcoming 4MOST survey \citep{2019Msngr.175....3D} will include a self-consistent determination of stellar abundances, including host stars, allowing to test our predictions.

\section{Summary and Conclusion}
\label{sec:summary}

We have presented here 1D simulations of inward drifting pebbles that can form planetesimals at the water ice line. In particular we focused here on the difference of the planetesimal formation efficiency in discs with low dust-to-gas ratios with varying C/O ratios. We thus present the first set of simulations where planetesimal formation is directly influenced by the composition of material rather than just by the overall amount of available solids.

We confirm previous simulations \citep{2024A&A...683A.118A} that find a lower limit in the overall dust-to-gas ratio for the formation of planetesimals in line with the drop of the occurrence rates of super-Earths \citep{2024AJ....168..128B}. We also confirm that disc with overall higher levels of turbulence seem to be more efficient to form planetesimals compared to discs with lower levels of turbulence. This is caused by the larger pebble sizes at low metallicities, which result in faster inward drift, reducing the time the disc can maintain a mid plane dust-to-gas ratio above unity at the water ice line, reducing the amount of planetesimals that can be formed under the assumption of efficient vertical settling \citep{Pinilla_2021}.

We find that lower C/O ratios - equivalent with higher water fractions - allow a more efficient formation planetesimals at the water ice live at the same overall dust-to-gas ratio. In particular, disc with dust-to-gas ratios below 0.6\% can only form planetesimals at the water ice line for low C/O ratios in our simulations. This implies that planets found around stars with low metallicity should preferentially be found around stars with low C/O ratios, testable with future detailed observations. Our simulations thus open a pathway to understand if the chemical composition of planet forming material influences the formation of planets.

\begin{acknowledgements}
K. O. Xenos thanks the Observatoire de la Côte d’Azur and the Erasmus+ program for making this joint research work possible. We thank the anonymous referee for the valuable comments that helped us to improve the overall quality of the paper.

\end{acknowledgements}

\bibliographystyle{aa}
\bibliography{Stellar.bib}

\begin{thebibliography}{79}
\expandafter\ifx\csname natexlab\endcsname\relax\def\natexlab#1{#1}\fi

\bibitem[{{Adibekyan} {et~al.}(2021){Adibekyan}, {Dorn}, {Sousa}, {Santos}, {Bitsch}, {Israelian}, {Mordasini}, {Barros}, {Delgado Mena}, {Demangeon}, {Faria}, {Figueira}, {Hakobyan}, {Oshagh}, {Soares}, {Kunitomo}, {Takeda}, {Jofr{\'e}}, {Petrucci}, \& {Martioli}}]{2021Sci...374..330A}
{Adibekyan}, V., {Dorn}, C., {Sousa}, S.~G., {et~al.} 2021, Science, 374, 330

\bibitem[{{Adibekyan} {et~al.}(2012){Adibekyan}, {Santos}, {Sousa}, {Israelian}, {Delgado Mena}, {Gonz{\'a}lez Hern{\'a}ndez}, {Mayor}, {Lovis}, \& {Udry}}]{2012A&A...543A..89A}
{Adibekyan}, V.~Z., {Santos}, N.~C., {Sousa}, S.~G., {et~al.} 2012, \aap, 543, A89

\bibitem[{{Andama} {et~al.}(2024){Andama}, {Mah}, \& {Bitsch}}]{2024A&A...683A.118A}
{Andama}, G., {Mah}, J., \& {Bitsch}, B. 2024, \aap, 683, A118

\bibitem[{Asplund {et~al.}(2009)Asplund, Grevesse, Sauval, \& Scott}]{2009ARA&A..47..481A}
Asplund, M., Grevesse, N., Sauval, A.~J., \& Scott, P. 2009, ARAA, 47, pp.481

\bibitem[{{Ataiee} {et~al.}(2018){Ataiee}, {Baruteau}, {Alibert}, \& {Benz}}]{2018A&A...615A.110A}
{Ataiee}, S., {Baruteau}, C., {Alibert}, Y., \& {Benz}, W. 2018, \aap, 615, A110

\bibitem[{{Bae} {et~al.}(2023){Bae}, {Isella}, {Zhu}, {Martin}, {Okuzumi}, \& {Suriano}}]{2023ASPC..534..423B}
{Bae}, J., {Isella}, A., {Zhu}, Z., {et~al.} 2023, in Astronomical Society of the Pacific Conference Series, Vol. 534, Protostars and Planets VII, ed. S.~{Inutsuka}, Y.~{Aikawa}, T.~{Muto}, K.~{Tomida}, \& M.~{Tamura}, 423

\bibitem[{{Batygin} \& {Morbidelli}(2023)}]{2023NatAs.tmp...10B}
{Batygin}, K. \& {Morbidelli}, A. 2023, Nature Astronomy

\bibitem[{{Bergner} {et~al.}(2024){Bergner}, {Sturm}, {Piacentino}, {McClure}, {{\"O}berg}, {Boogert}, {Dartois}, {Drozdovskaya}, {Fraser}, {Harsono}, {Ioppolo}, {Law}, {Lis}, {McGuire}, {Melnick}, {Noble}, {Palumbo}, {Pendleton}, {Perotti}, {Qasim}, {Rocha}, \& {van Dishoeck}}]{2024ApJ...975..166B}
{Bergner}, J.~B., {Sturm}, J.~A., {Piacentino}, E.~L., {et~al.} 2024, \apj, 975, 166

\bibitem[{Birnstiel {et~al.}(2012)Birnstiel, Klahr, \& Ercolano}]{2012A&A...539A.148B}
Birnstiel, T., Klahr, H., \& Ercolano, B. 2012, A\&A, 539, id.A148

\bibitem[{{Bitsch} \& {Battistini}(2020)}]{2020A&A...633A..10B}
{Bitsch}, B. \& {Battistini}, C. 2020, \aap, 633, A10

\bibitem[{{Bitsch} \& {Izidoro}(2023)}]{2023A&A...674A.178B}
{Bitsch}, B. \& {Izidoro}, A. 2023, \aap, 674, A178

\bibitem[{{Bitsch} \& {Izidoro}(2024)}]{2024A&A...692A.246B}
{Bitsch}, B. \& {Izidoro}, A. 2024, \aap, 692, A246

\bibitem[{Bitsch \& Johansen(2016)}]{2016A&A...590A.101B}
Bitsch, B. \& Johansen, A. 2016, A\&A, 590, id.A101

\bibitem[{{Bitsch} \& {Mah}(2023)}]{2023A&A...679A..11B}
{Bitsch}, B. \& {Mah}, J. 2023, \aap, 679, A11

\bibitem[{{Bitsch} {et~al.}(2018){Bitsch}, {Morbidelli}, {Johansen}, {Lega}, {Lambrechts}, \& {Crida}}]{2018arXiv180102341B}
{Bitsch}, B., {Morbidelli}, A., {Johansen}, A., {et~al.} 2018, A\&A, 612, id.A30

\bibitem[{{Boley} {et~al.}(2024){Boley}, {Christiansen}, {Zink}, {Hardegree-Ullman}, {Lee}, {Hopkins}, {Wang}, {Fernandes}, {Bergsten}, \& {Bhure}}]{2024AJ....168..128B}
{Boley}, K.~M., {Christiansen}, J.~L., {Zink}, J., {et~al.} 2024, \aj, 168, 128

\bibitem[{Brauer {et~al.}(2008)Brauer, Dullemond, \& Henning}]{2008A&A...480..859B}
Brauer, F., Dullemond, C., \& Henning, T. 2008, A\&A, 480, pp.859

\bibitem[{{Brewer} {et~al.}(2018){Brewer}, {Wang}, {Fischer}, \& {Foreman-Mackey}}]{2018ApJ...867L...3B}
{Brewer}, J.~M., {Wang}, S., {Fischer}, D.~A., \& {Foreman-Mackey}, D. 2018, \apjl, 867, L3

\bibitem[{Buchhave {et~al.}(2014)Buchhave, Bizzarro, Latham, Sasselov, Cochran, Endl, Isaacson, Juncher, \& Marcy}]{2014Natur.509..593B}
Buchhave, L.~A., Bizzarro, M., Latham, D.~W., {et~al.} 2014, Nature, 509, pp.593

\bibitem[{{Buchhave} {et~al.}(2012){Buchhave}, {Latham}, {Johansen}, {Bizzarro}, {Torres}, {Rowe}, {Batalha}, {Borucki}, {Brugamyer}, {Caldwell}, {Bryson}, {Ciardi}, {Cochran}, {Endl}, {Esquerdo}, {Ford}, {Geary}, {Gilliland}, {Hansen}, {Isaacson}, {Laird}, {Lucas}, {Marcy}, {Morse}, {Robertson}, {Shporer}, {Stefanik}, {Still}, \& {Quinn}}]{2012Natur.486..375B}
{Buchhave}, L.~A., {Latham}, D.~W., {Johansen}, A., {et~al.} 2012, \nat, 486, 375

\bibitem[{{Buder} {et~al.}(2018){Buder}, {Asplund}, {Duong}, {Kos}, {Lind}, {Ness}, {Sharma}, {Bland -Hawthorn}, {Casey}, {de Silva}, {D'Orazi}, {Freeman}, {Lewis}, {Lin}, {Martell}, {Schlesinger}, {Simpson}, {Zucker}, {Zwitter}, {Amarsi}, {Anguiano}, {Carollo}, {Casagrande}, {{\v C}otar}, {Cottrell}, {da Costa}, {Gao}, {Hayden}, {Horner}, {Ireland}, {Kafle}, {Munari}, {Nataf}, {Nordlander}, {Stello}, {Ting}, {Traven}, {Watson}, {Wittenmyer}, {Wyse}, {Yong}, {Zinn}, {{\v Z}erjal}, \& {Galah Collaboration}}]{2018MNRAS.478.4513B}
{Buder}, S., {Asplund}, M., {Duong}, L., {et~al.} 2018, \mnras, 478, 4513

\bibitem[{{Buder} {et~al.}(2021){Buder}, {Sharma}, {Kos}, {Amarsi}, {Nordlander}, {Lind}, {Martell}, {Asplund}, {Bland-Hawthorn}, {Casey}, {de Silva}, {D'Orazi}, {Freeman}, {Hayden}, {Lewis}, {Lin}, {Schlesinger}, {Simpson}, {Stello}, {Zucker}, {Zwitter}, {Beeson}, {Buck}, {Casagrande}, {Clark}, {{\v C}otar}, {da Costa}, {de Grijs}, {Feuillet}, {Horner}, {Kafle}, {Khanna}, {Kobayashi}, {Liu}, {Montet}, {Nandakumar}, {Nataf}, {Ness}, {Spina}, {Tepper-Garc\'{\i}a}, {Ting}, {Traven}, {Vogrin{\v c}i{\v c}}, {Wittenmyer}, {Wyse}, {{\v Z}erjal}, \& {Galah Collaboration}}]{2021MNRAS.506..150B}
{Buder}, S., {Sharma}, S., {Kos}, J., {et~al.} 2021, \mnras, 506, 150

\bibitem[{{Burbidge} {et~al.}(1957){Burbidge}, {Burbidge}, {Fowler}, \& {Hoyle}}]{1957RvMP...29..547B}
{Burbidge}, E.~M., {Burbidge}, G.~R., {Fowler}, W.~A., \& {Hoyle}, F. 1957, Reviews of Modern Physics, 29, 547

\bibitem[{Burke \& Brown(2010)}]{burke2010}
Burke, D.~J. \& Brown, W.~A. 2010, Phys. Chem. Chem. Phys., 12, 5947

\bibitem[{{Cabral, N.} {et~al.}(2023){Cabral, N.}, {Guilbert-Lepoutre, A.}, {Bitsch, B.}, {Lagarde, N.}, \& {Diakite, S.}}]{2023arXiv230105034C}
{Cabral, N.}, {Guilbert-Lepoutre, A.}, {Bitsch, B.}, {Lagarde, N.}, \& {Diakite, S.} 2023, A\&A, 673, A117

\bibitem[{{Carrera} {et~al.}(2017){Carrera}, {Gorti}, {Johansen}, \& {Davies}}]{2017ApJ...839...16C}
{Carrera}, D., {Gorti}, U., {Johansen}, A., \& {Davies}, M.~B. 2017, \apj, 839, 16

\bibitem[{Carrera {et~al.}(2015)Carrera, Johansen, \& Davies}]{2015A&A...579A..43C}
Carrera, D., Johansen, A., \& Davies, M.~B. 2015, A\&A, 579, id.A43

\bibitem[{{de Jong} {et~al.}(2019){de Jong}, {Agertz}, {Berbel}, {Aird}, {Alexander}, {Amarsi}, {Anders}, {Andrae}, {Ansarinejad}, {Ansorge}, {Antilogus}, {Anwand-Heerwart}, {Arentsen}, {Arnadottir}, {Asplund}, {Auger}, {Azais}, {Baade}, {Baker}, {Baker}, {Balbinot}, {Baldry}, {Banerji}, {Barden}, {Barklem}, {Barth{\'e}l{\'e}my-Mazot}, {Battistini}, {Bauer}, {Bell}, {Bellido-Tirado}, {Bellstedt}, {Belokurov}, {Bensby}, {Bergemann}, {Bestenlehner}, {Bielby}, {Bilicki}, {Blake}, {Bland-Hawthorn}, {Boeche}, {Boland}, {Boller}, {Bongard}, {Bongiorno}, {Bonifacio}, {Boudon}, {Brooks}, {Brown}, {Brown}, {Br{\"u}ggen}, {Brynnel}, {Brzeski}, {Buchert}, {Buschkamp}, {Caffau}, {Caillier}, {Carrick}, {Casagrande}, {Case}, {Casey}, {Cesarini}, {Cescutti}, {Chapuis}, {Chiappini}, {Childress}, {Christlieb}, {Church}, {Cioni}, {Cluver}, {Colless}, {Collett}, {Comparat}, {Cooper}, {Couch}, {Courbin}, {Croom}, {Croton}, {Daguis{\'e}}, {Dalton}, {Davies}, {Davis}, {de Laverny}, {Deason}, {Dionies}, {Disseau}, {Doel},
  {D{\"o}scher}, {Driver}, {Dwelly}, {Eckert}, {Edge}, {Edvardsson}, {Youssoufi}, {Elhaddad}, {Enke}, {Erfanianfar}, {Farrell}, {Fechner}, {Feiz}, {Feltzing}, {Ferreras}, {Feuerstein}, {Feuillet}, {Finoguenov}, {Ford}, {Fotopoulou}, {Fouesneau}, {Frenk}, {Frey}, {Gaessler}, {Geier}, {Gentile Fusillo}, {Gerhard}, {Giannantonio}, {Giannone}, {Gibson}, {Gillingham}, {Gonz{\'a}lez-Fern{\'a}ndez}, {Gonzalez-Solares}, {Gottloeber}, {Gould}, {Grebel}, {Gueguen}, {Guiglion}, {Haehnelt}, {Hahn}, {Hansen}, {Hartman}, {Hauptner}, {Hawkins}, {Haynes}, {Haynes}, {Heiter}, {Helmi}, {Aguayo}, {Hewett}, {Hinton}, {Hobbs}, {Hoenig}, {Hofman}, {Hook}, {Hopgood}, {Hopkins}, {Hourihane}, {Howes}, {Howlett}, {Huet}, {Irwin}, {Iwert}, {Jablonka}, {Jahn}, {Jahnke}, {Jarno}, {Jin}, {Jofre}, {Johl}, {Jones}, {J{\"o}nsson}, {Jordan}, {Karovicova}, {Khalatyan}, {Kelz}, {Kennicutt}, {King}, {Kitaura}, {Klar}, {Klauser}, {Kneib}, {Koch}, {Koposov}, {Kordopatis}, {Korn}, {Kosmalski}, {Kotak}, {Kovalev}, {Kreckel}, {Kripak}, {Krumpe},
  {Kuijken}, {Kunder}, {Kushniruk}, {Lam}, {Lamer}, {Laurent}, {Lawrence}, {Lehmitz}, {Lemasle}, {Lewis}, {Li}, {Lidman}, {Lind}, {Liske}, {Lizon}, {Loveday}, {Ludwig}, {McDermid}, {Maguire}, {Mainieri}, {Mali}, \& {Mandel}}]{2019Msngr.175....3D}
{de Jong}, R.~S., {Agertz}, O., {Berbel}, A.~A., {et~al.} 2019, The Messenger, 175, 3

\bibitem[{{Delgado Mena} {et~al.}(2021){Delgado Mena}, {Adibekyan}, {Santos}, {Tsantaki}, {Gonz{\'a}lez Hern{\'a}ndez}, {Sousa}, \& {Bertr{\'a}n de Lis}}]{2021A&A...655A..99D}
{Delgado Mena}, E., {Adibekyan}, V., {Santos}, N.~C., {et~al.} 2021, \aap, 655, A99

\bibitem[{{Dr{\c a}{\.z}kowska} \& {Alibert}(2017)}]{2017A&A...608A..92D}
{Dr{\c a}{\.z}kowska}, J. \& {Alibert}, Y. 2017, \aap, 608, A92

\bibitem[{{Dr{\k{a}}{\.z}kowska} {et~al.}(2023){Dr{\k{a}}{\.z}kowska}, {Bitsch}, {Lambrechts}, {Mulders}, {Harsono}, {Vazan}, {Liu}, {Ormel}, {Kretke}, \& {Morbidelli}}]{2023ASPC..534..717D}
{Dr{\k{a}}{\.z}kowska}, J., {Bitsch}, B., {Lambrechts}, M., {et~al.} 2023, in Astronomical Society of the Pacific Conference Series, Vol. 534, Protostars and Planets VII, ed. S.~{Inutsuka}, Y.~{Aikawa}, T.~{Muto}, K.~{Tomida}, \& M.~{Tamura}, 717

\bibitem[{Fischer \& Valenti(2005)}]{2005ApJ...622.1102F}
Fischer, D.~A. \& Valenti, J. 2005, ApJ, 622, pp. 1102

\bibitem[{Fressin {et~al.}(2013)Fressin, Torres, Charbonneau, Bryson, Christiansen, Dressing, Jenkins, Walkowicz, \& Batalha}]{2013ApJ...766...81F}
Fressin, F., Torres, G., Charbonneau, D., {et~al.} 2013, ApJ, 766, id.81

\bibitem[{{Gail} \& {Trieloff}(2017)}]{2017A&A...606A..16G}
{Gail}, H.-P. \& {Trieloff}, M. 2017, \aap, 606, A16

\bibitem[{{Gibb} {et~al.}(2004){Gibb}, {Whittet}, {Boogert}, \& {Tielens}}]{2004ApJS..151...35G}
{Gibb}, E.~L., {Whittet}, D.~C.~B., {Boogert}, A.~C.~A., \& {Tielens}, A.~G.~G.~M. 2004, \apjs, 151, 35

\bibitem[{{Gillon} {et~al.}(2017){Gillon}, {Triaud}, {Demory}, {Jehin}, {Agol}, {Deck}, {Lederer}, {de Wit}, {Burdanov}, {Ingalls}, {Bolmont}, {Leconte}, {Raymond}, {Selsis}, {Turbet}, {Barkaoui}, {Burgasser}, {Burleigh}, {Carey}, {Chaushev}, {Copperwheat}, {Delrez}, {Fernandes}, {Holdsworth}, {Kotze}, {Van Grootel}, {Almleaky}, {Benkhaldoun}, {Magain}, \& {Queloz}}]{2017Natur.542..456G}
{Gillon}, M., {Triaud}, A. H.~M.~J., {Demory}, B.-O., {et~al.} 2017, \nat, 542, 456

\bibitem[{{Guerra-Alvarado} {et~al.}(2025){Guerra-Alvarado}, {van der Marel}, {Williams}, {Pinilla}, {Mulders}, {Lambrechts}, \& {Sanchez}}]{2025A&A...696A.232G}
{Guerra-Alvarado}, O.~M., {van der Marel}, N., {Williams}, J.~P., {et~al.} 2025, \aap, 696, A232

\bibitem[{Gundlach \& Blum(2015)}]{2015ApJ...798...34G}
Gundlach, B. \& Blum, J. 2015, ApJ, 798, id. 34

\bibitem[{{Hayashi}(1981)}]{1981PThPS..70...35H}
{Hayashi}, C. 1981, Progress of Theoretical Physics Supplement, 70, pp.35

\bibitem[{{Hsieh} {et~al.}(2025){Hsieh}, {Arce}, {Jos{\'e} Maureira}, {Pineda}, {Segura-Cox}, {Mardones}, {Dunham}, {Li}, \& {Offner}}]{2025arXiv250411577H}
{Hsieh}, C.-H., {Arce}, H.~G., {Jos{\'e} Maureira}, M., {et~al.} 2025, arXiv e-prints, arXiv:2504.11577

\bibitem[{{Hsieh} {et~al.}(2024){Hsieh}, {Arce}, {Maureira}, {Pineda}, {Segura-Cox}, {Mardones}, {Dunham}, \& {Arun}}]{2024ApJ...973..138H}
{Hsieh}, C.-H., {Arce}, H.~G., {Maureira}, M.~J., {et~al.} 2024, \apj, 973, 138

\bibitem[{{Izidoro} {et~al.}(2021{\natexlab{a}}){Izidoro}, {Bitsch}, {Raymond}, {Johansen}, {Morbidelli}, {Lambrechts}, \& {Jacobson}}]{2019arXiv190208772I}
{Izidoro}, A., {Bitsch}, B., {Raymond}, S.~N., {et~al.} 2021{\natexlab{a}}, \aap, 650, A152

\bibitem[{{Izidoro} {et~al.}(2021{\natexlab{b}}){Izidoro}, {Dasgupta}, {Raymond}, {Deienno}, {Bitsch}, \& {Isella}}]{2022NatAs...6..357I}
{Izidoro}, A., {Dasgupta}, R., {Raymond}, S.~N., {et~al.} 2021{\natexlab{b}}, Nature Astronomy, 6, 357

\bibitem[{Izidoro {et~al.}(2017)Izidoro, Ogihara, Raymond, Morbidelli, Pierens, Bitsch, Cossou, \& Hersant}]{2017MNRAS.470.1750I}
Izidoro, A., Ogihara, M., Raymond, S.~N., {et~al.} 2017, MNRAS, 470, pp. 1750

\bibitem[{Johansen {et~al.}(2015)Johansen, {Mac Low}, Lacerda, \& Bizzarro}]{Johansen2015}
Johansen, A., {Mac Low}, M.~M., Lacerda, P., \& Bizzarro, M. 2015, Science Advances, Vol.1, id. 1500109

\bibitem[{{Johansen} {et~al.}(2007){Johansen}, {Oishi}, {Mac Low}, {Klahr}, {Henning}, \& {Youdin}}]{2007Natur.448.1022J}
{Johansen}, A., {Oishi}, J.~S., {Mac Low}, M.~M., {et~al.} 2007, Nature, 448, pp. 1022

\bibitem[{{Johansen} \& {Youdin}(2007)}]{2007ApJ...662..627J}
{Johansen}, A. \& {Youdin}, A. 2007, ApJ, 662, pp. 627

\bibitem[{{Johnson} {et~al.}(2010){Johnson}, {Aller}, {Howard}, \& {Crepp}}]{J2010}
{Johnson}, J.~A., {Aller}, K.~M., {Howard}, A.~W., \& {Crepp}, J.~R. 2010, PASP, 122, 905

\bibitem[{Lambrechts {et~al.}(2014)Lambrechts, Johansen, \& Morbidelli}]{2014A&A...572A..35L}
Lambrechts, M., Johansen, A., \& Morbidelli, A. 2014, A\&A, 572, id. A35

\bibitem[{{Lambrechts} {et~al.}(2019){Lambrechts}, {Morbidelli}, {Jacobson}, {Johansen}, {Bitsch}, {Izidoro}, \& {Raymond}}]{2019A&A...627A..83L}
{Lambrechts}, M., {Morbidelli}, A., {Jacobson}, S.~A., {et~al.} 2019, \aap, 627, A83

\bibitem[{{Lenz} {et~al.}(2019){Lenz}, {Klahr}, \& {Birnstiel}}]{2019ApJ...874...36L}
{Lenz}, C.~T., {Klahr}, H., \& {Birnstiel}, T. 2019, \apj, 874, 36

\bibitem[{{Li} \& {Youdin}(2021)}]{2021ApJ...919..107L}
{Li}, R. \& {Youdin}, A.~N. 2021, \apj, 919, 107

\bibitem[{Lodders(2003)}]{2003ApJ...591.1220L}
Lodders, K. 2003, ApJ, 591, pp. 1220

\bibitem[{Madhusudhan {et~al.}(2014)Madhusudhan, Amin, \& Kennedy}]{2014ApJ...794L..12M}
Madhusudhan, N., Amin, M.~A., \& Kennedy, G.~M. 2014, ApJL, 794, L12

\bibitem[{{Mathis} {et~al.}(1977){Mathis}, {Rumpl}, \& {Nordsieck}}]{1977ApJ...217..425M}
{Mathis}, J.~S., {Rumpl}, W., \& {Nordsieck}, K.~H. 1977, \apj, 217, 425

\bibitem[{Mayor \& Queloz(1995)}]{1995Natur.378..355M}
Mayor, M. \& Queloz, D. 1995, Nature, 378, pp.355

\bibitem[{{Millholland} \& {Winn}(2021)}]{2021ApJ...920L..34M}
{Millholland}, S.~C. \& {Winn}, J.~N. 2021, \apjl, 920, L34

\bibitem[{{Mills} {et~al.}(2016){Mills}, {Fabrycky}, {Migaszewski}, {Ford}, {Petigura}, \& {Isaacson}}]{2016Natur.533..509M}
{Mills}, S.~M., {Fabrycky}, D.~C., {Migaszewski}, C., {et~al.} 2016, \nat, 533, 509

\bibitem[{{Morbidelli}(2020)}]{2020A&A...638A...1M}
{Morbidelli}, A. 2020, \aap, 638, A1

\bibitem[{{Mulders} {et~al.}(2018){Mulders}, {Pascucci}, {Apai}, \& {Ciesla}}]{2018AJ....156...24M}
{Mulders}, G.~D., {Pascucci}, I., {Apai}, D., \& {Ciesla}, F.~J. 2018, \aj, 156, 24

\bibitem[{{M{\"u}ller} {et~al.}(2021){M{\"u}ller}, {Savvidou}, \& {Bitsch}}]{2021A&A...650A.185M}
{M{\"u}ller}, J., {Savvidou}, S., \& {Bitsch}, B. 2021, \aap, 650, A185

\bibitem[{{Mumma} \& {Charnley}(2011)}]{2011ARA&A..49..471M}
{Mumma}, M.~J. \& {Charnley}, S.~B. 2011, \araa, 49, 471

\bibitem[{{Musiolik} \& {Wurm}(2019)}]{2019ApJ...873...58M}
{Musiolik}, G. \& {Wurm}, G. 2019, \apj, 873, 58

\bibitem[{{Narang} {et~al.}(2018){Narang}, {Manoj}, {Furlan}, {Mordasini}, {Henning}, {Mathew}, {Banyal}, \& {Sivarani}}]{2018AJ....156..221N}
{Narang}, M., {Manoj}, P., {Furlan}, E., {et~al.} 2018, \aj, 156, 221

\bibitem[{{Ohashi} {et~al.}(2023){Ohashi}, {Tobin}, {J{\o}rgensen}, {Takakuwa}, {Sheehan}, {Aikawa}, {Li}, {Looney}, {Williams}, {Aso}, {Sharma}, {Sai}, {Yamato}, {Lee}, {Tomida}, {Yen}, {Encalada}, {Flores}, {Gavino}, {Kido}, {Han}, {Lin}, {Narayanan}, {Phuong}, {Santamar{\'\i}a-Miranda}, {Thieme}, {van't Hoff}, {de Gregorio-Monsalvo}, {Koch}, {Kwon}, {Lai}, {Lee}, {Plunkett}, {Saigo}, {Hirano}, {Lam}, \& {Mori}}]{2023ApJ...951....8O}
{Ohashi}, N., {Tobin}, J.~J., {J{\o}rgensen}, J.~K., {et~al.} 2023, \apj, 951, 8

\bibitem[{{Petigura} {et~al.}(2018){Petigura}, {Marcy}, {Winn}, {Weiss}, {Fulton}, {Howard}, {Sinukoff}, {Isaacson}, {Morton}, \& {Johnson}}]{2018AJ....155...89P}
{Petigura}, E.~A., {Marcy}, G.~W., {Winn}, J.~N., {et~al.} 2018, \aj, 155, 89

\bibitem[{Pinilla {et~al.}(2021)Pinilla, Lenz, \& Stammler}]{Pinilla_2021}
Pinilla, P., Lenz, C.~T., \& Stammler, S.~M. 2021, \aap, 645, A70

\bibitem[{{S{\'a}ndor} {et~al.}(2024){S{\'a}ndor}, {Guilera}, {Reg{\'a}ly}, \& {Lyra}}]{2024A&A...686A..78S}
{S{\'a}ndor}, Z., {Guilera}, O.~M., {Reg{\'a}ly}, Z., \& {Lyra}, W. 2024, \aap, 686, A78

\bibitem[{Santos {et~al.}(2004)Santos, Israelia, \& Mayor}]{2004A&A...415.1153S}
Santos, N.~C., Israelia, G., \& Mayor, M. 2004, A\&A, 415, p.1153

\bibitem[{{Schneider} \& {Bitsch}(2021)}]{2021A&A...654A..71S}
{Schneider}, A.~D. \& {Bitsch}, B. 2021, \aap, 654, A71

\bibitem[{{Shakura} \& {Sunyaev}(1973)}]{1973A&A....24..337S}
{Shakura}, N.~I. \& {Sunyaev}, R.~A. 1973, \aap, 24, 337

\bibitem[{{Shibata} \& {Izidoro}(2025)}]{2025ApJ...979L..23S}
{Shibata}, S. \& {Izidoro}, A. 2025, \apjl, 979, L23

\bibitem[{Simon {et~al.}(2016)Simon, Armitage, Li, \& Youdin}]{2016ApJ...822...55S}
Simon, J., Armitage, P.~J., Li, R., \& Youdin, A. 2016, ApJ, 822, id. 55

\bibitem[{{Voelkel} {et~al.}(2022){Voelkel}, {Klahr}, {Mordasini}, \& {Emsenhuber}}]{2022A&A...666A..90V}
{Voelkel}, O., {Klahr}, H., {Mordasini}, C., \& {Emsenhuber}, A. 2022, \aap, 666, A90

\bibitem[{Weidenschilling(1977)}]{1977MNRAS.180...57W}
Weidenschilling, S.~J. 1977, MNRAS, 180, p.57

\bibitem[{{Weidenschilling}(1977)}]{1977Ap&SS..51..153W}
{Weidenschilling}, S.~J. 1977, \apss, 51, 153

\bibitem[{{Weiss} {et~al.}(2018{\natexlab{a}}){Weiss}, {Isaacson}, {Marcy}, {Howard}, {Petigura}, {Fulton}, {Winn}, {Hirsch}, {Sinukoff}, {Rowe}, \& {California Kepler Survey}}]{2018AJ....156..254W}
{Weiss}, L.~M., {Isaacson}, H.~T., {Marcy}, G.~W., {et~al.} 2018{\natexlab{a}}, \aj, 156, 254

\bibitem[{{Weiss} {et~al.}(2018{\natexlab{b}}){Weiss}, {Marcy}, {Petigura}, {Fulton}, {Howard}, {Winn}, {Isaacson}, {Morton}, {Hirsch}, {Sinukoff}, {Cumming}, {Hebb}, \& {Cargile}}]{2018AJ....155...48W}
{Weiss}, L.~M., {Marcy}, G.~W., {Petigura}, E.~A., {et~al.} 2018{\natexlab{b}}, \aj, 155, 48

\bibitem[{{Zubko} {et~al.}(2004){Zubko}, {Dwek}, \& {Arendt}}]{2004ApJS..152..211Z}
{Zubko}, V., {Dwek}, E., \& {Arendt}, R.~G. 2004, \apjs, 152, 211

\end{thebibliography}

\newpage

\appendix
\section{Solar Metallicity}
\label{app:solar}

\begin{figure}
  \centering
  \includegraphics[width=\hsize]{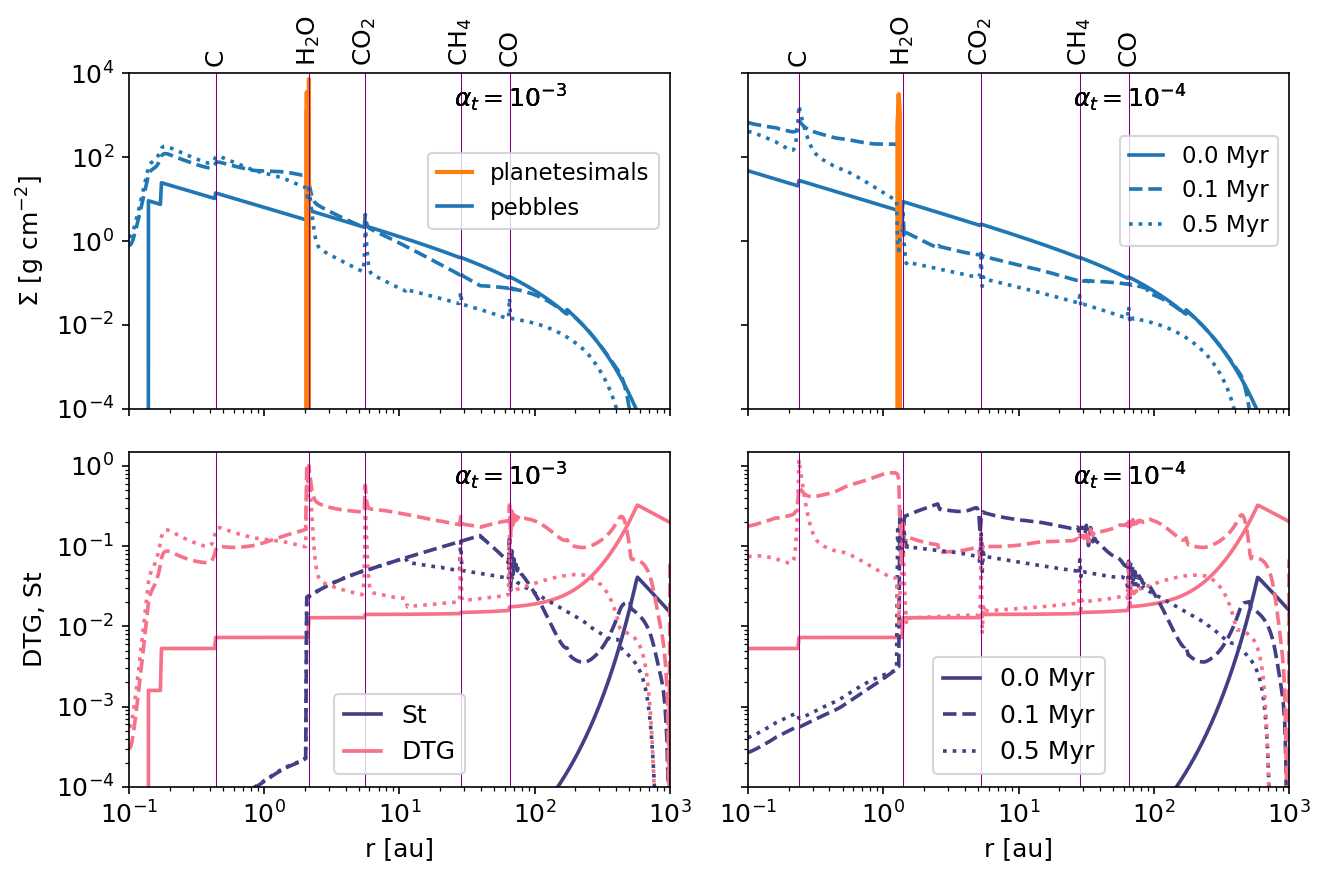}
      \caption{Solid surface density of pebbles and planetesimals (top) in a disc with [Fe/H] = 0.0 and the median C/O ratio for $\alpha=10^{-3}$ (left) and $\alpha=10^{-4}$ (right) as function of time. The bottom panels depict the Stokes numbers of the pebbles as well as the mid-plane dust-to-gas ratio.}
         \label{fig:solar}
  \end{figure}

Fig.~\ref{fig:solar} shows the solid surface density of pebbles and planetesimals at solar metallicity for different viscosities as well as the corresponding Stokes numbers and mid-plane dust-to-gas ratios. We use here the solar C/O ratio. Independently of the disc's viscosity, planetesimal formation at the water ice line is very efficient. Our here displayed results at correspond to the setup of \citet{2024A&A...683A.118A} and our simulations recover their results. Our simulations show a planetesimal formation efficiency of around $\approx 40$\% at high viscosity and $\approx 7$\% at low viscosity (see Fig.~\ref{fig:massesfull} and Fig.~\ref{fig:massesfulloh}).

\section{Water content}
\label{app:water}

\begin{figure}
  \centering
  \includegraphics[width=\hsize]{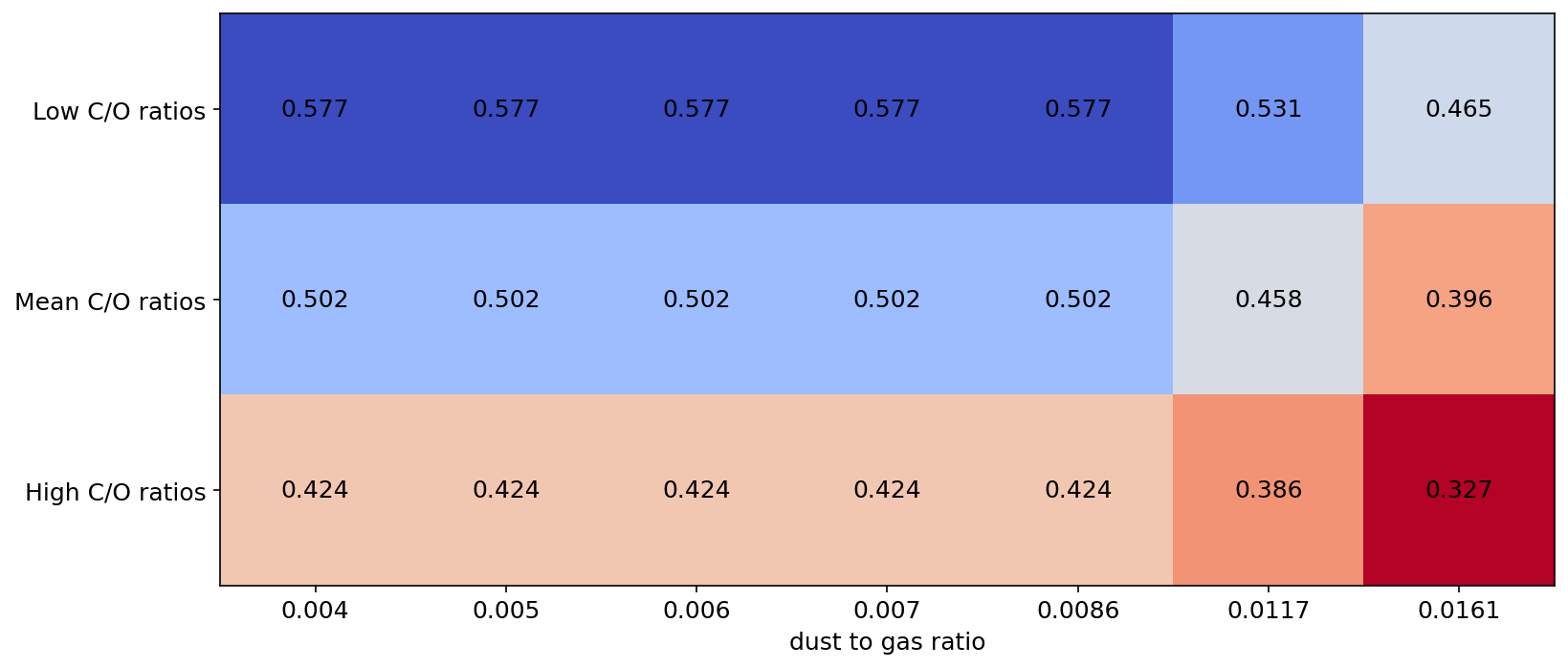}
    \includegraphics[width=\hsize]{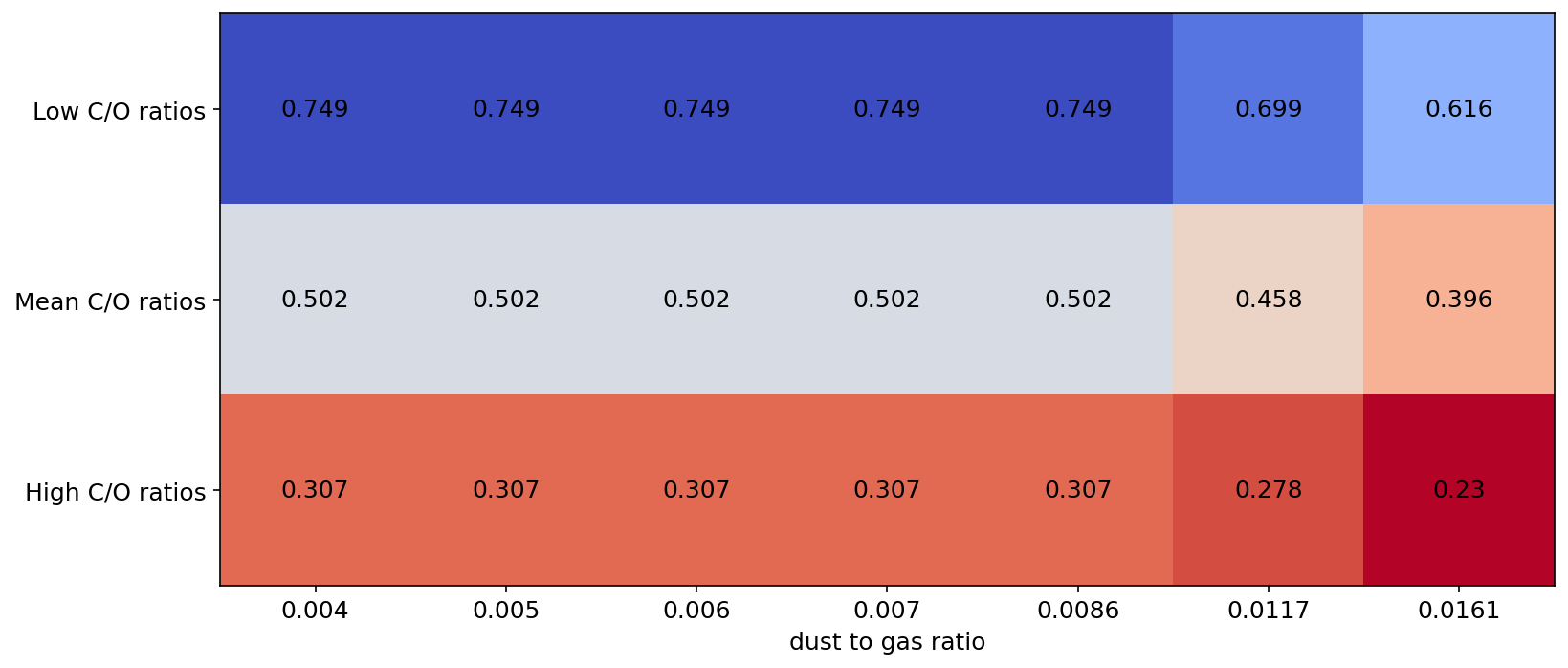}
      \caption{Water mass fraction in the solids at 140K - so just outside the water ice line - at the start of our simulations. The top plot depicts the water ice fraction when the C/O ratio is changed by varying the carbon value, while the bottom plot shows the C/O ratio values if the oxygen abundances is varied.}
         \label{fig:water}
  \end{figure}

We show in Fig.~\ref{fig:water} the water content of the solids in the protoplanetary disc at 140K, so just outside of the water ice line at the beginning of our simulations. We show the water mass fraction at the beginning of our simulations, as this gives the best understanding how the C/O ratio (see table~\ref{getCH})changes the water ice fraction. As time progresses, pebbles move inwards and release their water content into the gas phase. However, the outward diffusing water vapour can condense again and increase the water ice fraction at the ice line. This process - as discussed in the main paper - depends on the water ice fraction itself.

The water ice fraction depends clearly on the C/O ratio, but also on how the C/O ratio is manipulated. In our work, the water ice fraction is calculated by first distributing the elements into all carbon and oxygen bearing species, before the leftover oxygen is used to produce water. For low C/O ratios, where the carbon abundance is reduce, only the oxygen that is stored in CO and CO$_2$ is released to form water. However, the majority of carbon is actually contained in CH$_4$ and refractory carbon grains that harbour no oxygen. By reducing the carbon abundance for an overall low C/O ratio, only an tiny amount of oxygen is release, resulting in an increase of the water ice mass fraction by only 7\%. Reciprocally, the high C/O ratio then only results in a reducing of the water mass fraction by a similarly small amount.

On the other hand, if the C/O ratio is manipulated by increasing the oxygen fraction, a larger change in the water content can be achieved. This is caused by the fact that the total dust mass stays constant (constant dust-to-gas ratio) and an increase in oxygen results in a decrease of all other elements. This releases additional oxygen due to the less abundant Mg$_2$SiO$_4$ and MgSiO$_3$ - in addition to the less abundance CO and CO$_2$ - to form water. Consequently, the water ice fraction is increased by nearly 25\%, if the C/O ratio changes from 0.44 to 0.19. Reciprocally, an increase of the C/O ratio to 0.69 results in a decrease of the water fraction by nearly 20\%.

\label{LastPage}
\end{document}